\documentclass[12pt,fleqn]{article}

\usepackage{latexsym}

\usepackage[english]{babel}

\usepackage{amsmath,amssymb,bm}
\usepackage{indentfirst}

\usepackage[dvips]{graphicx}

\begin{document}

\title{The Shadow of  Generalized Kerr Black Holes with Exotic Matter}
\author{Vassil K. Tinchev\thanks{E-mail: tintschev@phys.uni-sofia.bg, tintschev@gmail.com}\\
{\footnotesize  Department of Theoretical Physics,
                Faculty of Physics, Sofia University,}\\
{\footnotesize  5 James Bourchier Boulevard, Sofia~1164, Bulgaria }\\}

\date{}

\maketitle

\begin{abstract}
We explore the shadow of certain class generalized Kerr black holes, which are non vacuum solutions of the Einstein equations with exotic matter. The images depend on the angular momentum of the compact object, the characteristic parameter $l$ of the model, and the inclination angle of the observer. The results are compared with the case of the Kerr black hole and the Kerr naked singularity.
\end{abstract}

The apparent shapes of compact relativistic objects contains information about them which can not be found by other ways \cite{FalckeEtAll}, \cite{Luminet}. Therefore the shadows of compact relativistic objects have been studied with major interest in the last years. Question about shadows of the Kerr-Newmann black holes' family have been explored in \cite{Bardeen}-\cite{bibl2}. The case of the black hole with  NUT-charge has been analysed in \cite{AbdujabbarovEtAll}. The studies \cite{AmarillaEiroa}-\cite{AmarillaEiroa1} investigate the black hole shadows in Einstein-Maxwell-dilaton gravity,  Chern-Simons modified gravity and braneworld gravity. The shadows of the Sen black hole, the wormhole, and the Kerr black hole pierced by cosmic string have been investigated in \cite{HiokiMiyamoto}-\cite{TinchevYazadjiev}. The shadow of Einstein-Maxwell-Dilaton-Axion black hole is presented in \cite{WeiLiu}. There are also results in higher dimensions as \cite{PapnoiEtAll}. Some interesting results about the shadows of Multi-Black Holes are presented in \cite{YumotoEtAll}.

Now we have a technology for experimental observation of the compact objects' shadows. It's the Event Horizon Telescope \cite{EHTelescope}, the orbital radio telescopes RadioAstron, Millimetron (\cite{Radioastron}, \cite{JohannsenEtAll}), and the X-ray interferometer MAXIM \cite{MAXIM}. These missions are expected to observe the apparent shape of the supermassive black hole at the center of our galaxy or those located at nearby galaxies \cite{JohannsenEtAll}, and it will be finished in the next few years. The theoretical models should be compared with the results of these experiments, which will alow to reject some of the models or will make it possible to distinguish between different types of compact objects. It also be conceivable to use the mentioned observations for detection of theoretically predicted objects and effects, which are not observed so far.

It is commonly accepted that the rotating black holes are describe by the Kerr solution of General relativity. This belief is justified by the fact that
General relativity is very successful theory of  gravity. However, there are observational phenomena, like the accelerated expansion of the Universe and dark matter,
which do not fit complete in the framework of General relativity and the standard model of particle physics. That is why new alternatives should be explored.
The existence of exotic matter in the form of dark energy or dark matter may have a serious influence on the structure of the black holes and their gravitational field
giving rise to new effects and phenomena. The presence of exotic matter in the interior and the exterior of the black holes can deform the usual Kerr solution
to new one with different spacetime geometry which in turn would effect the gravitational lensing of the black hole and its shadow in particular.
The aim of the current paper is to investigate the shadow of such  generalized Kerr black holes, and compare the results with the case of the usual Kerr black hole.
In modeling the generalized Kerr black holes we shall follow  simple principles. We require the spacetime metric of the generalized Kerr solution to admit  Killing tensor
in addition to the stationary and axisymmetric Killing vectors. This ensures the existence of third constant of motion, the sa-called Carter constant, which facilitates
significantly the study of the shadow. The deviation from the usual Kerr metric is required to be governed by only one function of the radial coordinate.
One of the simplest metrics that satisfies these requirements can be constructed from the Kerr metric by replacing the mass parameter in the Kerr solution by
an appropriate function of the radial coordinate, $M\to\mu(r)$. Similar approach for construction of generalized Kerr solutions  was also considered in \cite{Ghosh}.
Then the metric of the generalized Kerr black hole is given by

\begin{equation}\label{metric}
  ds^2=-\left(1-\frac{2\mu r}{\rho^{2}}\right)dt^2+\rho^{2}\left(\frac{dr^2}{\Delta}\ +
  d\theta^{2}\right)+\frac{\sin^2\theta}{\rho^{2}}\left(\Sigma^{2}d\varphi-4a\mu r\
  dt\right)d\varphi,
\end{equation}
where   $r$, $\theta$ and $\varphi$ are the Boyer-Lindquist coordinates and the usual definitions have been used for the metric functions $\Delta$, $\rho$ and $\Sigma$
\begin{equation}\label{metricFunc}
  \Delta\equiv r^2-2\mu r+a^2,\ \rho^2\equiv r^2+a^2\cos^2\theta,\ \Sigma^2\equiv
  \left(a^2+r^2\right)^{2}-a^2\Delta\sin^2\theta.
\end{equation}
The metric (\ref{metric}) reduces to the metric of Kerr when $\mu=M=const$.  Here we will consider the function  $\mu=Mexp(-l/r^2)$ where $l$  is a parameter.
This function does not effect the post-Newtonian regime of the Kerr metric up to third order and the parameter $l$ is unconstrained by current observational tests of  the gravitational theories. The metric (\ref{metric}) describes a rotating  asymptotically flat black hole with mass $M$ and angular momentum per unit of mass $a$. The black hole horizon is located at
$\Delta=0$. In order to check that the spacetime geometry is regular on the horizon and outside it we compute the   Kretschmann scalar:

\begin{equation}\label{curvInvKretschmann}
    \begin{array}{ll}
    K\equiv R^{\alpha\beta\gamma\delta}R_{\alpha\beta\gamma\delta}=\frac{4}{\rho^{12}}\left\{384r^{4}F^{2}
    -192r^{2}\rho^{2}F\left(2F+rF'\right)+\right.\\
    \\
    \left.\ \ \ \ \ \ \ \ \ \ \ \ \ \ \ \ \ \ \ \ \ \ \ \ +8\rho^{4}\left[7F^{2}+4r^{2}F'^{2}+2rF\left(8F'+rF''\right)\right]-\right.\\
    \\
    \left.\ \ \ \ \ \ \ \ \ \ \ \ \ \ \ \ \ \ \ \ \ \ \ \ -8\rho^{6}\left[3F'^{2}
    +F''\left(F+2rF'\right)\right]+\rho^{8}F''^{2}\right\},
    \end{array}
\end{equation}
where the function $F$ is defined by

\begin{eqnarray}
F(r)\equiv r\mu(r).
\end{eqnarray}
From the explicit expression (\ref{curvInvKretschmann}) one can see that the  Kretschmann scalar is regular everywhere on and outside the horizon.

Now let us consider the matter energy-momentum tensor associated with the metric (\ref{metric}). In the orthonormal basis $\omega^{a}_{\ \alpha}$ explicitly given by
\begin{equation}\label{omegaVecCov}
    \begin{array}{ll}
    \omega^{0}_{\ \alpha}=\frac{\sqrt{\Delta}}{\rho}\left(1,0,0,-a\sin^{2}\theta\right),\\
    \\
    \omega^{1}_{\ \alpha}=\frac{\rho}{\sqrt{\Delta}}\left(0,1,0,0\right),\\
    \\
    \omega^{2}_{\ \alpha}=\frac{\sin\theta}{\rho}\left(a,0,0,-r^{2}-a^{2}\right),\\
    \\
    \omega^{3}_{\ \alpha}=\rho\left(0,0,1,0\right),
    \end{array}
\end{equation}
with $g_{\alpha\beta}=\eta_{ab}\omega^{a}_{\ \alpha}\omega^{b}_{\ \beta}$  where $\eta_{ab}=diag(-1,1,1,1)$ is the Minkowski metric, the
components of the matter energy-momentum tensor are
\begin{equation}\label{Tmn}
    \begin{array}{ll}
    T_{ab}=\left(\frac{F-rF'}{8\pi\rho^{4}}\right)
    diag\left(-2,2,-2-\frac{\rho^{2}F''}{F-rF'},-2-\frac{\rho^{2}F''}{F-rF'}\right)=diag(\varrho,p_{1},p_{2},p_{3}).\\
    \end{array}
\end{equation}
Hence it is easy to see that all the energy conditions are violated and therefore our solution describes a rotating black hole in the presence
of exotic matter inside and outside the horizon. Outside the horizon the energy density and the pressure of the exotic matter falls off rapidly and the solution
tends to the Kerr solution.

\begin{figure}[h]
        \setlength{\tabcolsep}{ 0 pt }{\scriptsize\tt
        \begin{tabular}{ cc }
            \includegraphics[width=0.7\textwidth]{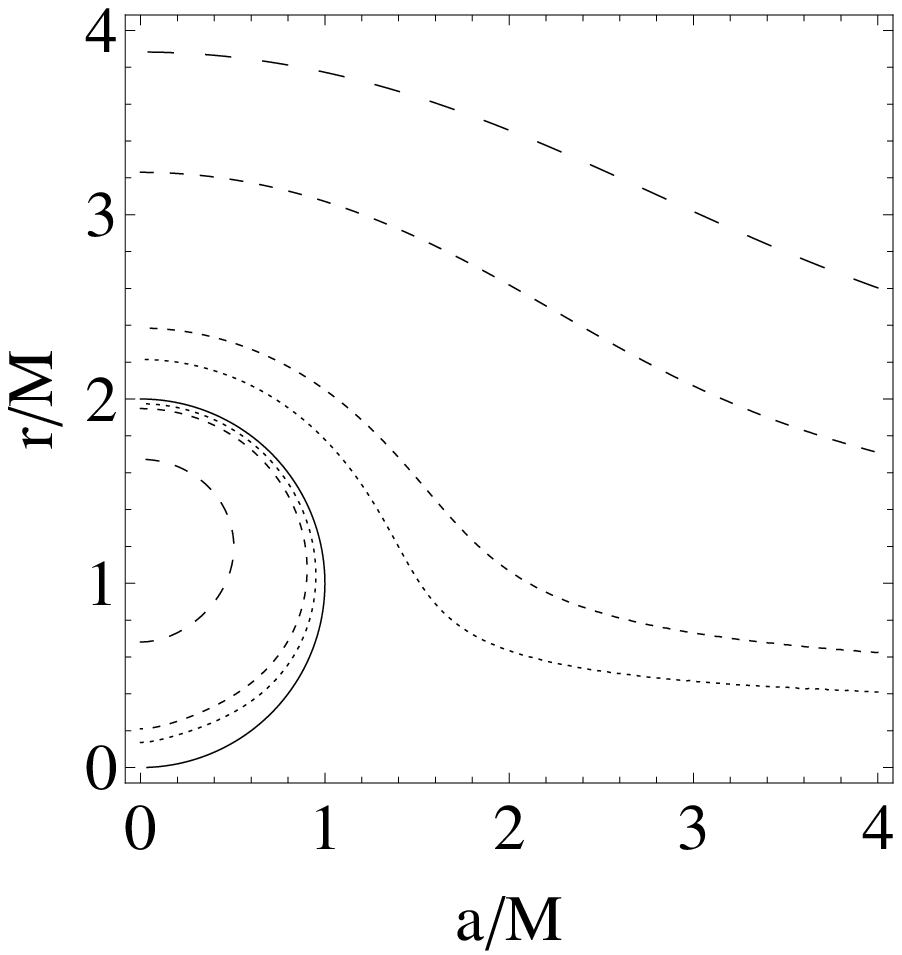} &
            \\
        \end{tabular}}
\caption{\footnotesize{The curves $\Delta(a,l,r)/M^{2}=0$ for some fixed values of the parameter $l/M^{2}$. From top right to bottom left $l/M^{2}$ is equal to $-10$, $-5$, $-1$, $-0.5$, $0$ (solid line), $0.05$, $0.1$, and $0.5$ respectively. } }
        \label{fig2}
\end{figure}

 In Fig. \ref{fig2} we present the curves $\Delta(a,l,r)/M^{2}=0$  for some fixed values of the parameter $l/M^{2}$. We see that for $l\geq 0$ it have an intervals about $a$ ($0\leq a<a_{m}(l)\leq M$) where exist two horizons of our metric as in the case of the pure Kerr black hole ($l=0$), where $a_{m}(0)=M$. For $l<0$ our black hole solution can violate the
Kerr bound and there can be black holes with $a>M$ which correspond to naked singularities in the case of the pure Kerr solution.

The  Hamilton-Jacobi equation determines the motion of test particles in our space-time,
\begin{equation}\label{HamiltonJacobiEq}
-2\frac{\partial S}{\partial \lambda}= g^{\alpha\beta}\frac{\partial
S}{\partial x^{\alpha}}\frac{\partial S}{\partial x^{\beta}},
\end{equation}
where $S$ is the particle action, $\lambda$ is the affine parameter
along the geodesics of the metric $g_{\alpha\beta}$. Two conserved quantities:- the energy of the particle $E$ and its angular momentum $L_{z}$ about the axis of symmetry, result from the stationary and axisymmetric nature of our space-time described by the metric (\ref{metric}). We also have another conserved quantity, namely the Carter constant, as in the case of pure Kerr space-time. This leads to the separability of the Hamilton-Jacobi equation, with a solution of the form
\begin{equation}\label{solHamiltonJacobiEq}
  S=\frac{1}{2}m^{2}\lambda-Et+L_{z}\varphi+S_{r}(r)+S_{\theta}(\theta),
\end{equation}
where $m$ is the mass of the test particle.  Using
(\ref{solHamiltonJacobiEq}) the Hamilton-Jacobi equation reduces to
the following equations for $S_{r}(r)$ and $S_{\theta}(\theta)$:
\begin{equation}\label{soleqrtheta1}
    \begin{array}{ll}
      \left(S'_{r}\right)^{2}=\frac{\left(a^{2}+r^{2}\right)^{2}}{\Delta^{2}}\ E^{2}
      +\frac{a^{2}}{\Delta^{2}}\ L_{z}^{2}-\frac{4a\mu r}{\Delta^{2}}\ EL_{z}-\frac{m^{2}r^{2}+{\cal K}}{\Delta}\equiv \frac{R(r)}{\Delta^2},\\
      \\
      \left(S'_{\theta}\right)^{2}={\cal K}-m^{2}a^{2}\cos^{2}\theta-E^{2}a^{2}\sin^{2}\theta-
      \frac{1}{\sin^{2}\theta}\ L_{z}^{2}\equiv
      \Theta(\theta),
    \end{array}
\end{equation}
where ${\cal K}$ is a Carter constant. Then \eqref{solHamiltonJacobiEq} takes the form
\begin{equation}\label{solHamiltonJacobiEqConcrete}
    S=\frac{1}{2}m^{2}\lambda-Et+L_{z}\varphi+\int \frac{1}{\Delta}\sqrt{R(r)}\ dr+\int \sqrt{\Theta(\theta)}\ d\theta.
\end{equation}
Using standard procedure we are able to find the null geodesics
(i.e. $m=0$) in our space-time, as follows:
\begin{equation}\label{eqGeodesics1AIsotropic}
    \begin{array}{ll}
        \rho^{2}\ \frac{dt}{d\lambda}=\frac{1}{\Delta}\left[\left(a^{2}+r^{2}\right)^{2}-
    2a\xi\mu r\right]-a^{2}\sin^{2}\theta,\\
        \\
        \rho^{2}\ \frac{dr}{d\lambda}=\pm\sqrt{\left(a^{2}+r^{2}\right)^{2}-
    a\xi\left(4\mu r-a\xi\right)-\Delta\eta}\equiv\pm\sqrt{R},\\
        \\
        \rho^{2}\ \frac{d\theta}{d\lambda}=\pm\sqrt{\eta-a^{2}\sin^{2}\theta-
      \frac{1}{\sin^{2}\theta}\ \xi^{2}}\equiv\pm\sqrt{\Theta},\\
        \\
        \rho^{2}\ \frac{d\varphi}{d\lambda}=\frac{\xi}{\sin^{2}\theta}+
    \frac{a}{\Delta}\left(2\mu r-a\xi\right),
    \end{array}
\end{equation}
with   $\xi\equiv L_{z}/E$ and $\eta\equiv {\cal K}/E^{2}$ being the impact
parameters. We have also redefined the affine parameter  $E\lambda
\to \lambda$.

There are generally two types of photon orbits; those falling into the black hole and those scattered to infinity away from the black hole. When observed from great distances, in relation to the black hole, it is possible only to see the scattered photons. Those photons captured by the black hole form the dark region at the centre. This dark region, observed on the luminous background, is the shadow of the black hole. The boundary of this shadow is the critical point that separates the escape orbit from the plunge orbit. We must we reformulate the problem as one-dimensional problem for a particle in an effective potential, in order to find the shadow boundary. This is achieved by rewriting the radial geodesic equation in the form
\begin{equation*}
  \left(\rho^{2}\
  \frac{dr}{d\lambda}\right)^{2}+U_{eff}(r)=0,
\end{equation*}
where $U_{eff}(r)=-R(r)$. It is clear from the above formulation that the critical orbit between escape and plunge, corresponds to the highest maximum of the effective potential.  It is also obvious that the orbit is both unstable and circular. Therefore, the conditions for the spherical form of the critical orbit that truly determines the boundary of the black hole shadow are as follows
\begin{eqnarray}\label{CONU}
U_{eff}=0, \;\; \frac{dU_{eff}}{dr}=0, \;\;
\frac{d^2U_{eff}}{dr^2}\le 0,
\end{eqnarray}
or, equivalently, $R=0$, $\frac{dR}{dr}=0$ and $\frac{d^2R}{dr^2}\ge
0$. There is in fact a parametric relation between the impact parameters that should be satisfied on the shadow boundary since the effective potential $U_{eff}$ (or equivalently $R$) depends on $r$ as well as  $\xi$ and $\eta$, the conditions (\ref{CONU}). Of course; the impact parameters should be such that $\Theta(\theta)\ge 0$, in addition to the  conditions above.

Taking into account the explicit form of the function $R$ in our
case, the solution to the conditions $R=0$ and $\frac{dR}{dr}=0$,
that also satisfies $\Theta(\theta)\ge 0$, is given by
\begin{equation}\label{solxieta2}
    \begin{array}{ll}
    \xi=\frac{1}{a(F'-r)}\left[(a^2+r^2)(F'+r)-4rF\right],\\
    \\
    \eta=\frac{2}{(F'-r)^{2}}\left[(a^2+r^2)(F'^{2}+r^{2})-4rFF'\right].
    \end{array}
\end{equation}
The condition  $\frac{d^2R}{dr^2}\ge 0$  leads to the following explicit  inequality
\begin{equation}\label{d2Rrsolxieta2}
    3r^2+a^2+\frac{(a^2+r^2)\left[r^{2}(2F''-1)-F'^{2}\right]-4rF(rF''-F')}{(F'-r)^{2}}\ge
    0.
\end{equation}
Equations (\ref{solxieta2}) and (\ref{d2Rrsolxieta2}) define the
boundary of the shadow in  parametric form. It is clear from the
derivation that the boundary of the shadow does not depend on the details of the emission
mechanisms, and is determined only by the space-time  metric.

In reality what is seen by an observer, is in fact the projection of the shadow on the sky. This should be defined as the plane that passes through the black hole and the line of sight. Therefore, it is more natural to present the shadow boundary in the so-called celestial coordinates $\alpha$ and $\beta$. The celestial coordinates are defined by \cite{bibl1}
\begin{equation}\label{limalpha}
    \alpha= \lim_{r\to\ \infty}\left(-r^{2}\sin\theta_{0}\frac{d\varphi}{dr}\right),
\end{equation}
\begin{equation}\label{limbeta}
  \beta=\lim_{r\to\ \infty}\left(r^{2}\frac{d\theta}{dr}\right),
\end{equation}
where the limit is taken along the null geodesics and $\theta_{0}$
is the inclination angle between the axis of rotation of the black
hole and the line of sight of the observer. From the definition of
the celestial coordinates and using the null geodesics equations
\eqref{eqGeodesics1AIsotropic}, in case when $\lim\limits_{r\to\ \infty}\mu(r)=const$, we get
\begin{equation}\label{alphabetaA}
    \begin{array}{ll}
  \alpha=-\frac{\xi}{\sin\theta_{0}}\ ,\\
  \\
  \beta=\pm\sqrt{\eta-a^{2}\sin^{2}\theta_{0}-\frac{\xi^{2}}{\sin^{2}\theta_{0}}}\ .
    \end{array}
\end{equation}
After substituting \eqref{solxieta2} in \eqref{alphabetaA}, we find
\begin{equation}\label{solalphabetaA2}
    \begin{array}{ll}
    \alpha=-\frac{1}{a(F'-r)\sin\theta_{0}}\left[(a^2+r^2)(F'+r)-4rF\right],\\
    \\
    \beta=\pm\sqrt{\frac{2\left[(a^2+r^2)(F'^{2}+r^{2})-4rFF'\right]}{(F'-r)^{2}}-a^{2}\sin^{2}\theta_{0}-
    \frac{\left[(a^2+r^2)(F'+r)-4rF\right]^{2}}{a^{2}(F'-r)^{2}\sin^{2}\theta_{0}}}.
    \end{array}
\end{equation}
These two equations give the boundary of the shadow in celestial coordinates in parametric form. We see that the equations (\ref{solalphabetaA2}) are not in the form of pure Kerr black hole. Therefore it will be easy to separate the shadows of the compact relativistic objects from this type, and the shadows of Kerr black holes.

\begin{figure}[h]
        \setlength{\tabcolsep}{ 0 pt }{\scriptsize\tt
        \begin{tabular}{ cccc }
            \includegraphics[width=0.25\textwidth]{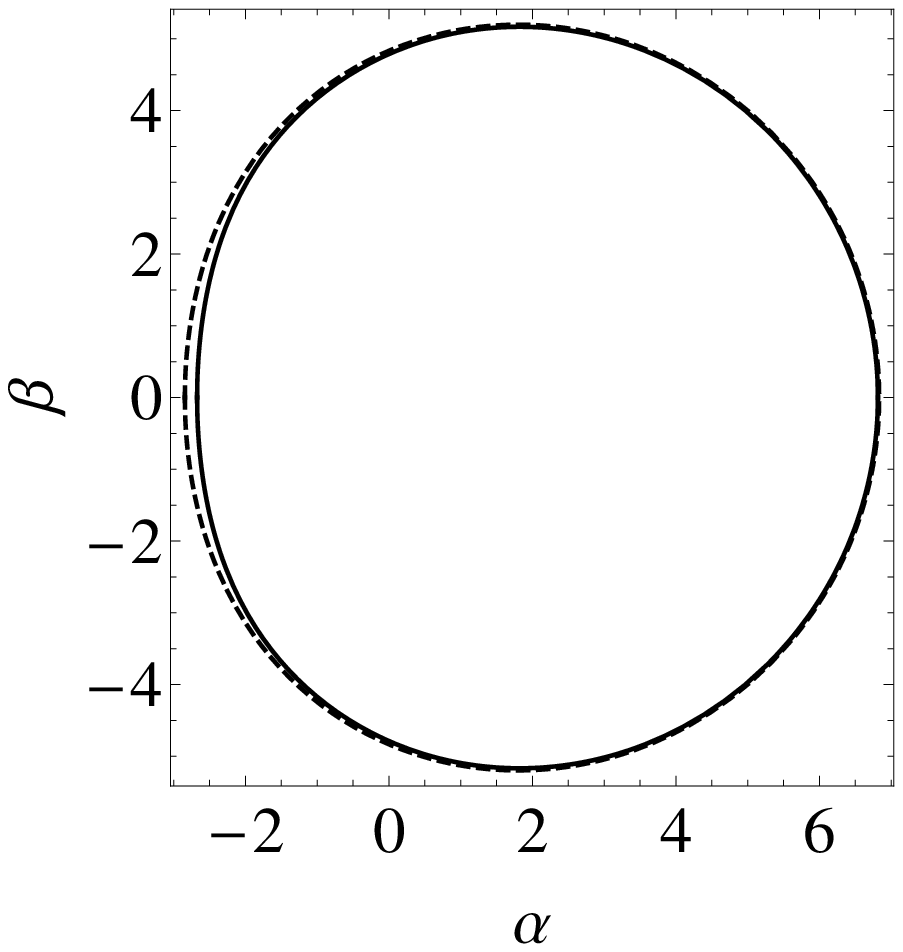} &
            \includegraphics[width=0.25\textwidth]{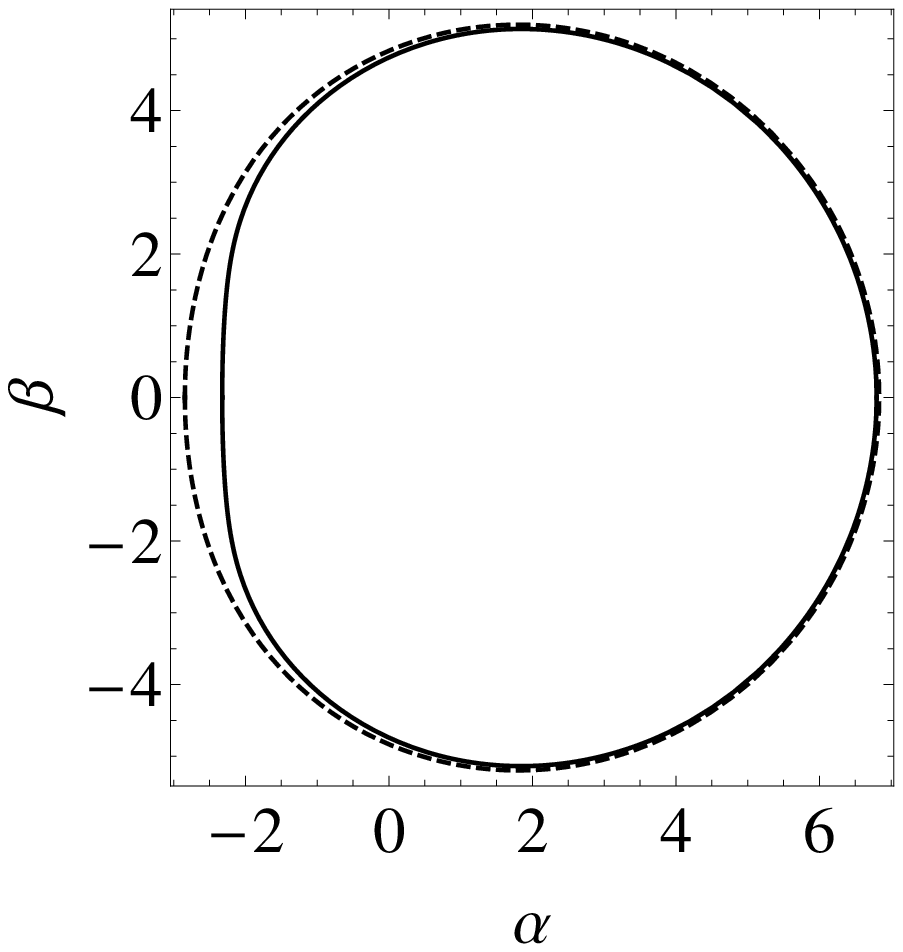} &
            \includegraphics[width=0.25\textwidth]{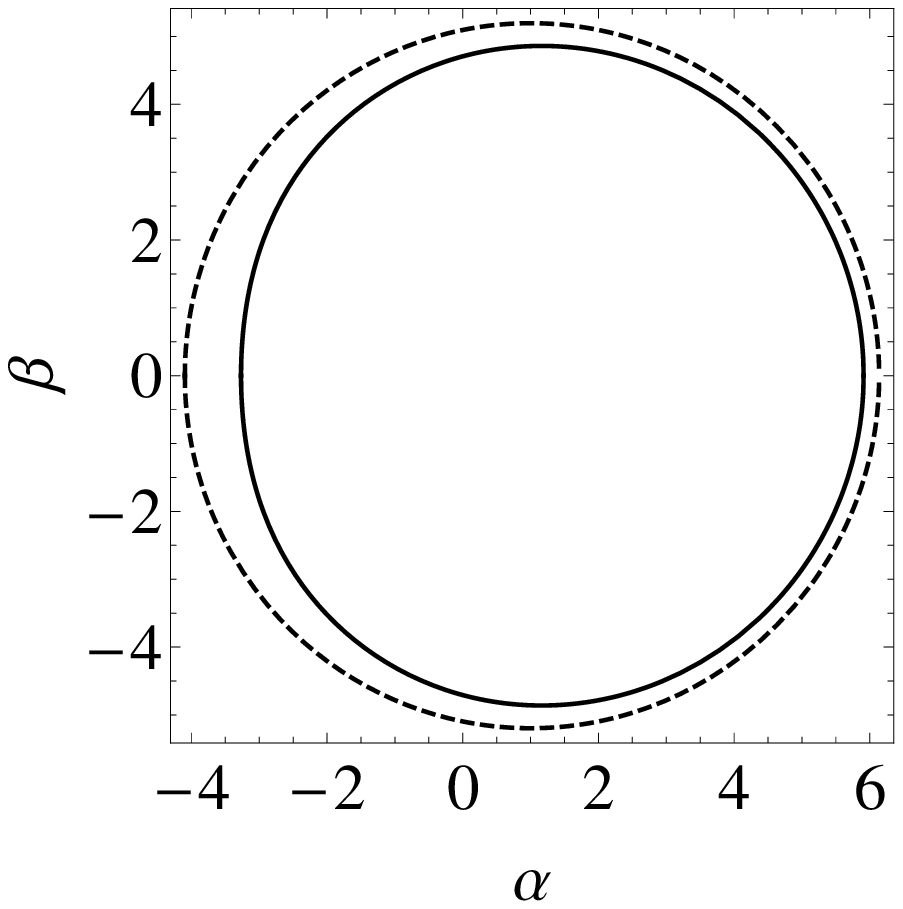} &
            \includegraphics[width=0.25\textwidth]{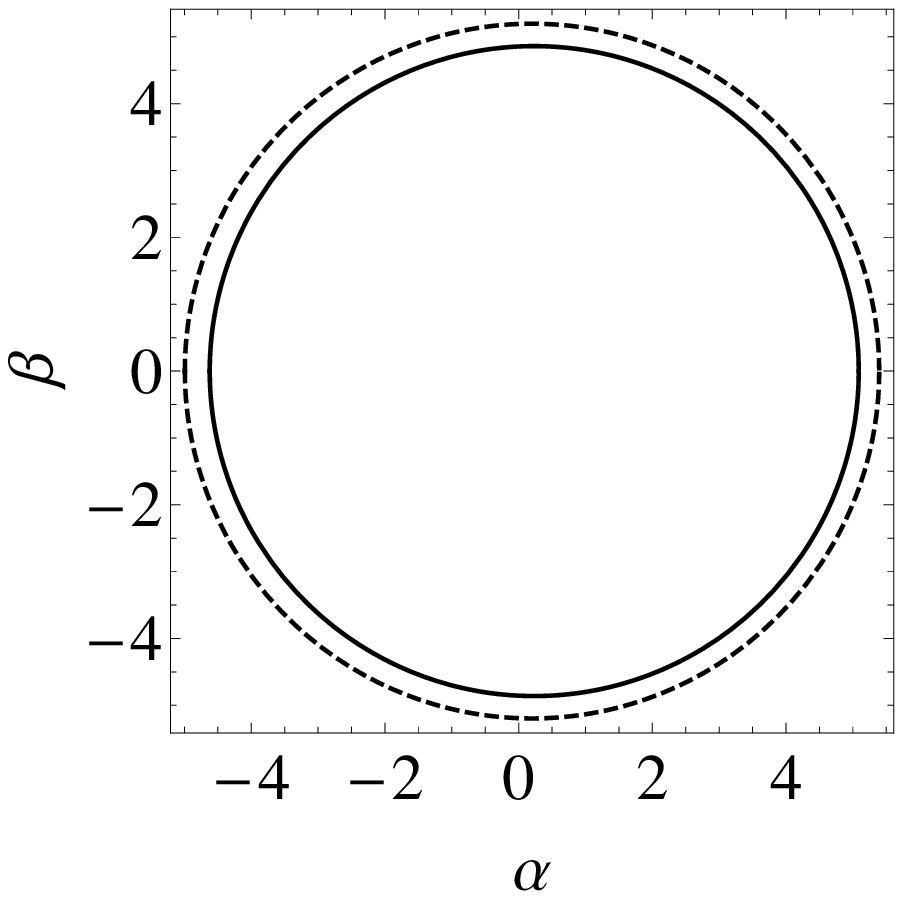} \\
            $a/M=0.9$, $l/M^{2}=0.05$;\  &
            $a/M=0.9$, $l/M^{2}=0.1$;\  &
            $a/M=0.5$, $l/M^{2}=0.5$;\  &
            $a/M=0.1$, $l/M^{2}=0.5$ \\
        \end{tabular}}
\caption{\footnotesize{The shadow of the generalized Kerr black hole in the model with function $\mu(r)=M e^{-l/r^{2}}$ (solid line) and the Kerr black hole (dashed line)
with inclination angle $\theta_{0}=\pi/2\ rad$ for different
values of parameters $a$ and $l$. The parameter $M$ of both solutions
is set equal to 1. The celestial coordinates $(\alpha,\beta)$ are
measured in the units of parameter $M$. } }
        \label{WS_a3}
\end{figure}

\begin{figure}[h]
        \setlength{\tabcolsep}{ 0 pt }{\scriptsize\tt
        \begin{tabular}{ cccc }
            \includegraphics[width=0.25\textwidth]{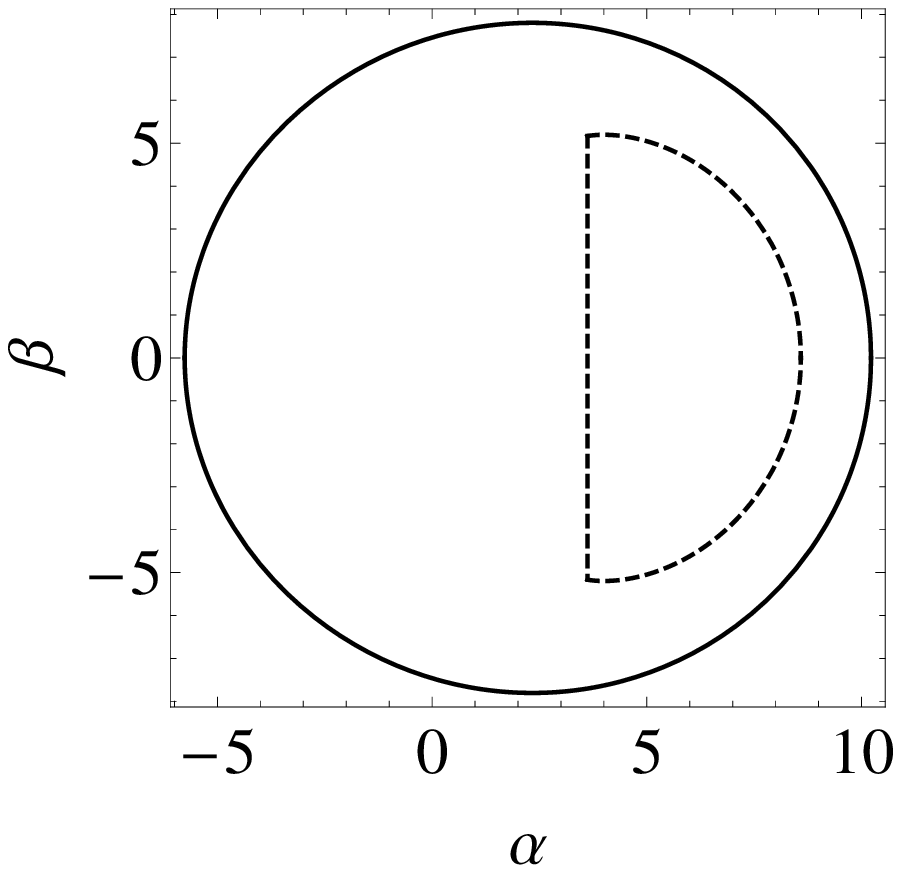} &
            \includegraphics[width=0.25\textwidth]{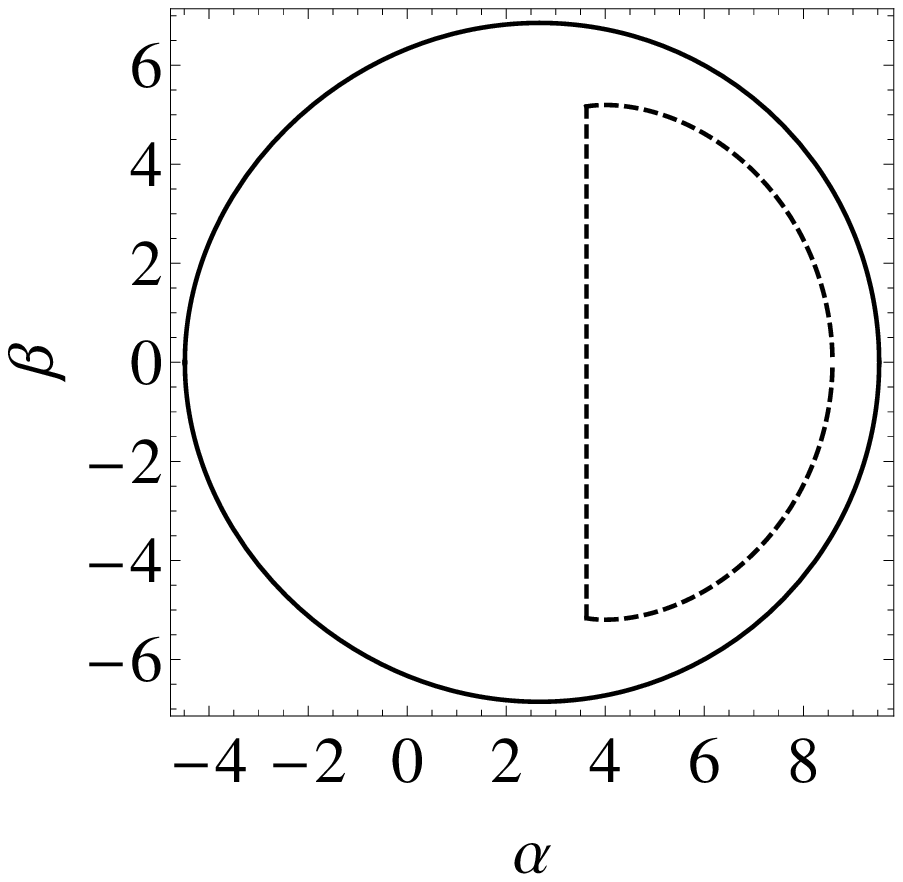} &
            \includegraphics[width=0.25\textwidth]{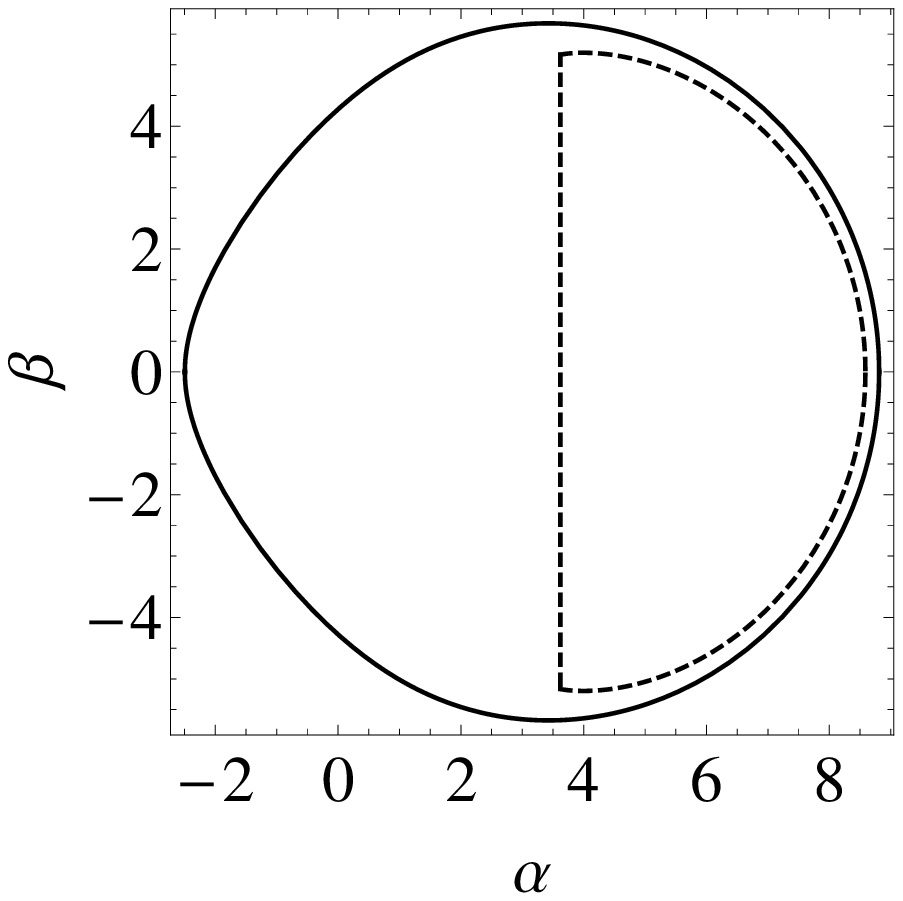} &
            \includegraphics[width=0.25\textwidth]{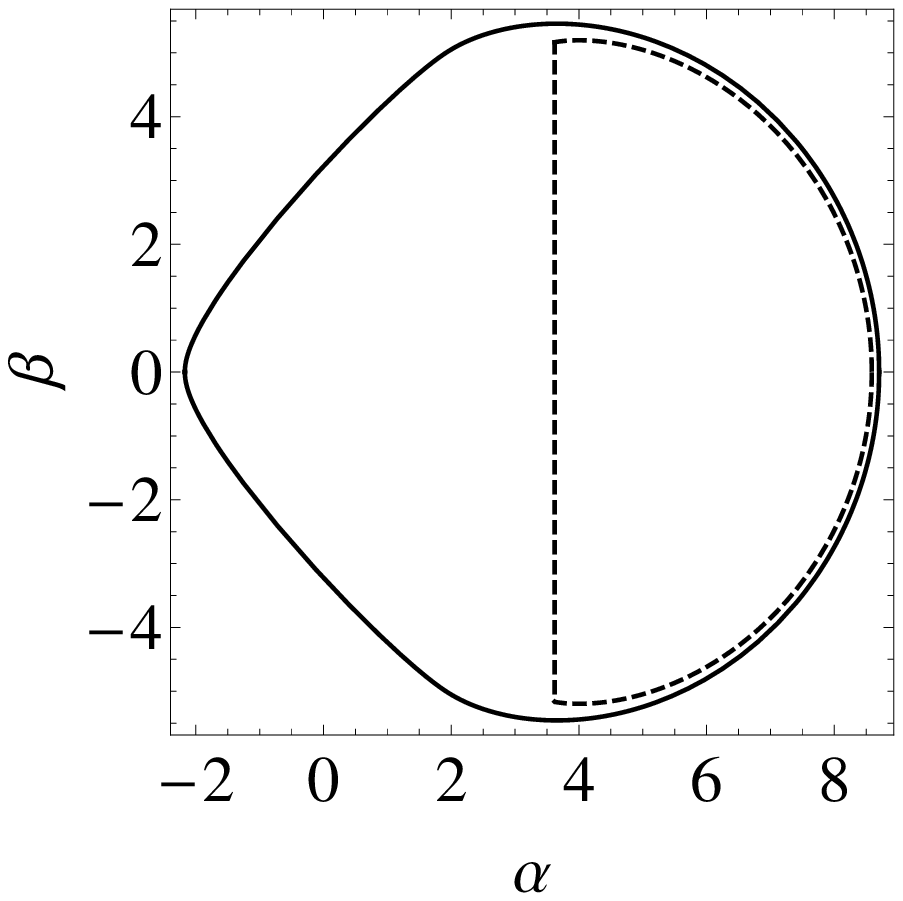} \\
            $a/M=2$, $l/M^{2}=-10$;\  &
            $a/M=2$, $l/M^{2}=-5$;\  &
            $a/M=2$, $l/M^{2}=-1$;\  &
            $a/M=2$, $l/M^{2}=-0.5$ \\
            \includegraphics[width=0.25\textwidth]{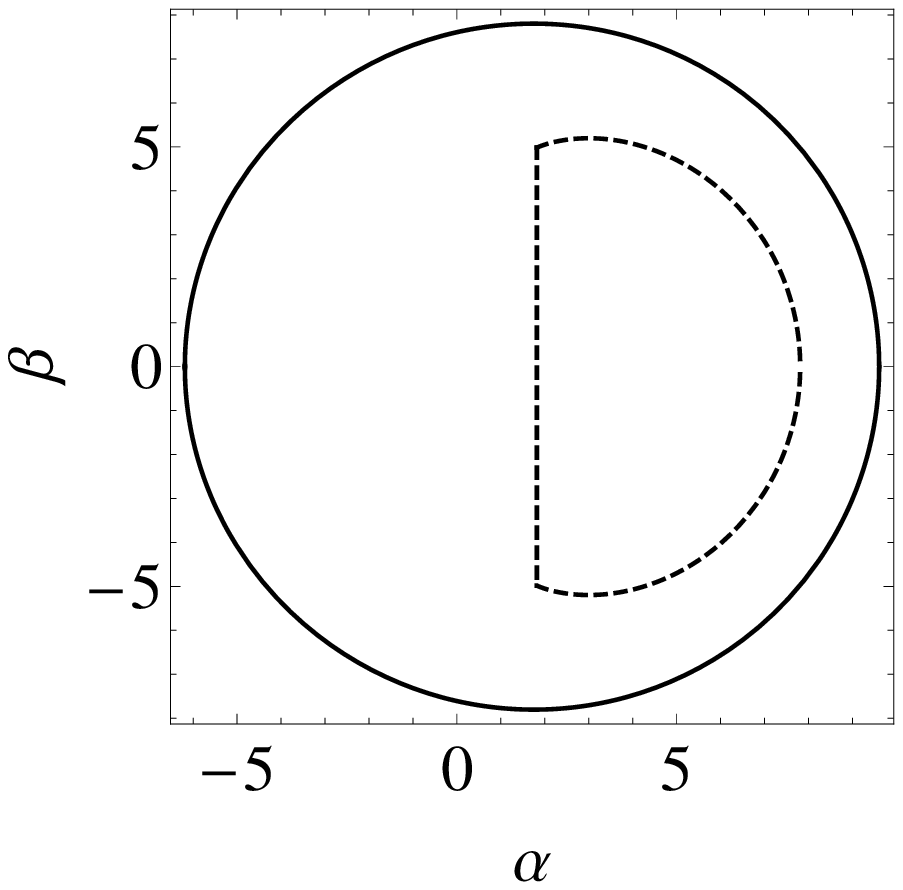} &
            \includegraphics[width=0.25\textwidth]{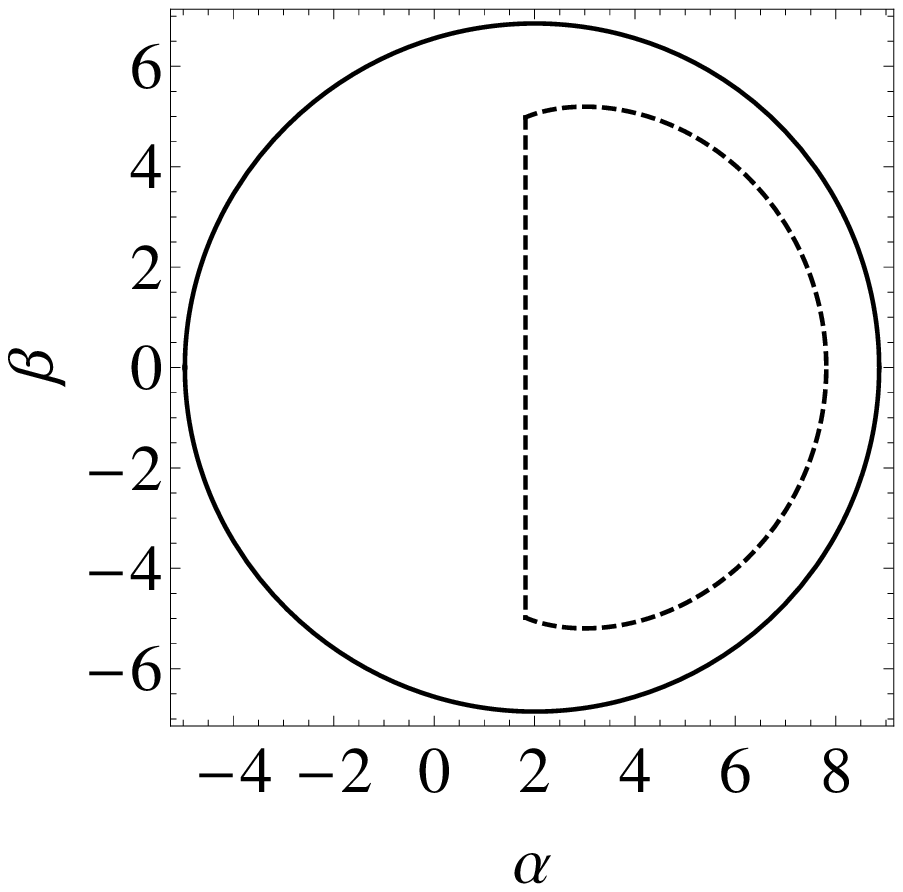} &
            \includegraphics[width=0.25\textwidth]{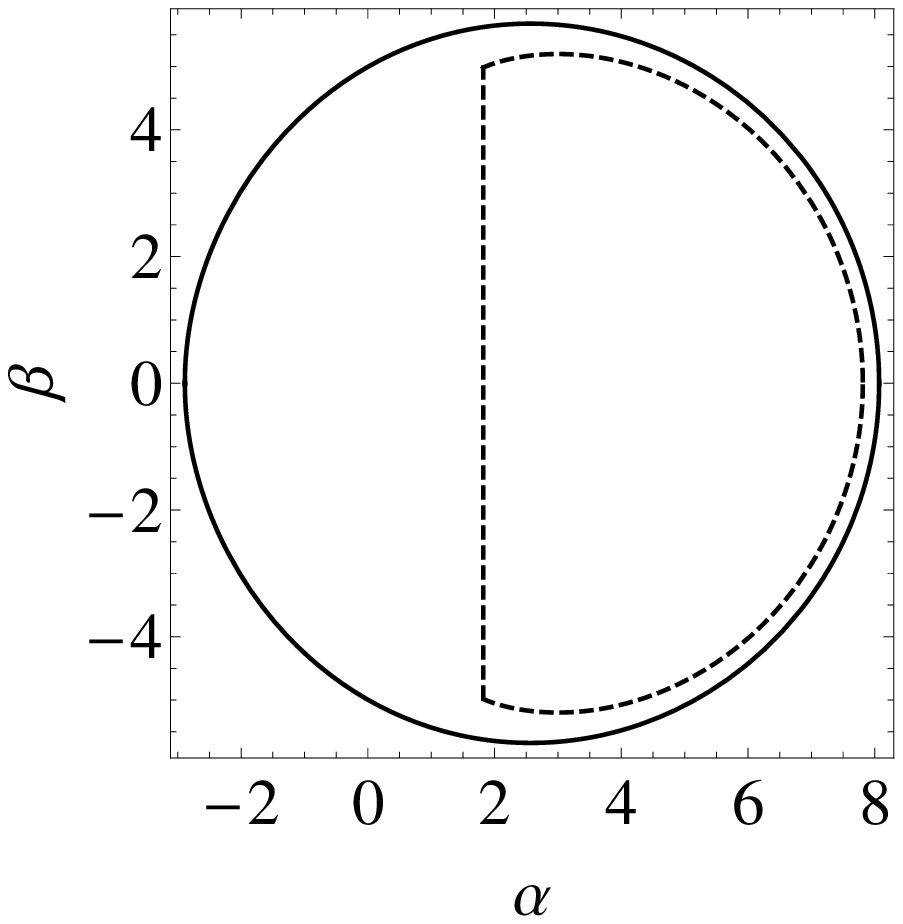} &
            \includegraphics[width=0.25\textwidth]{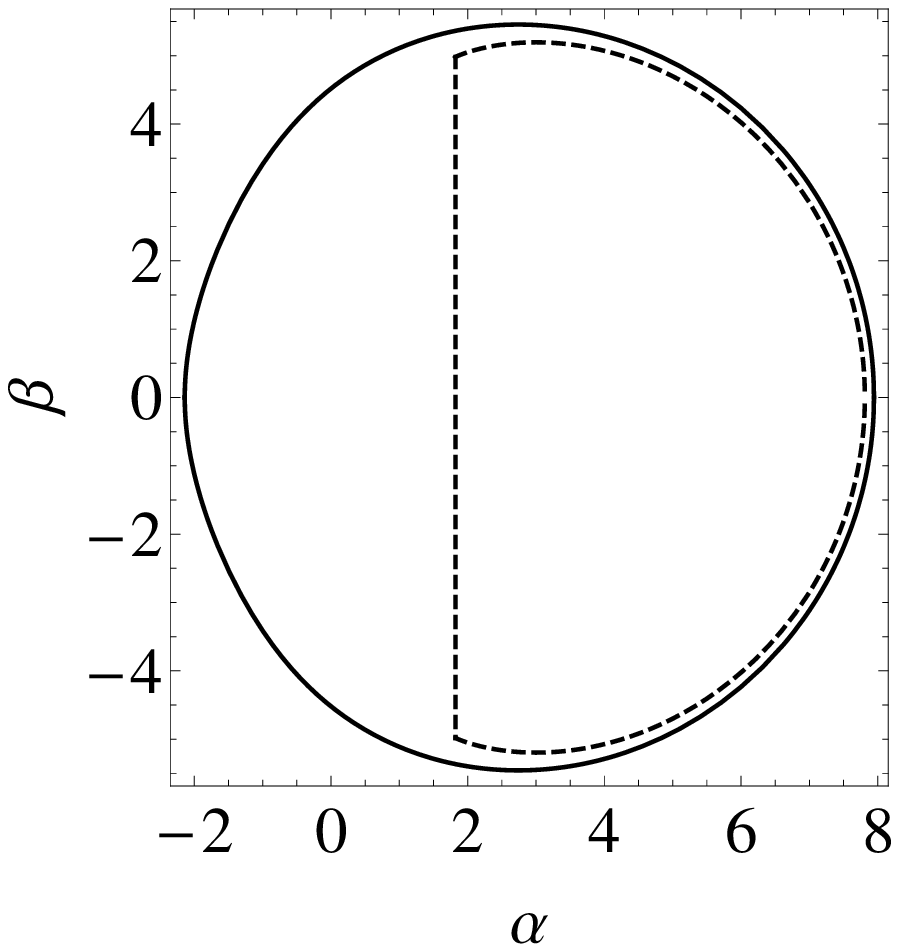} \\
            $a/M=1.5$, $l/M^{2}=-10$;\  &
            $a/M=1.5$, $l/M^{2}=-5$;\  &
            $a/M=1.5$, $l/M^{2}=-1$;\  &
            $a/M=1.5$, $l/M^{2}=-0.5$ \\
        \end{tabular}}
\caption{\footnotesize{The shadow of the generalized Kerr black hole in the model with function $\mu(r)=M e^{-l/r^{2}}$ (solid line) and the corresponding Kerr naked singularity (dashed line)
with inclination angle $\theta_{0}=\pi/2\ rad$ for different
values of parameters $a$ and $l$. The parameter $M$ of both solutions
is set equal to 1. The celestial coordinates $(\alpha,\beta)$ are
measured in the units of parameter $M$. } }
        \label{WS_a5minus}
\end{figure}

\begin{figure}[h]
        \setlength{\tabcolsep}{ 0 pt }{\scriptsize\tt
        \begin{tabular}{ cccc }
            \includegraphics[width=0.25\textwidth]{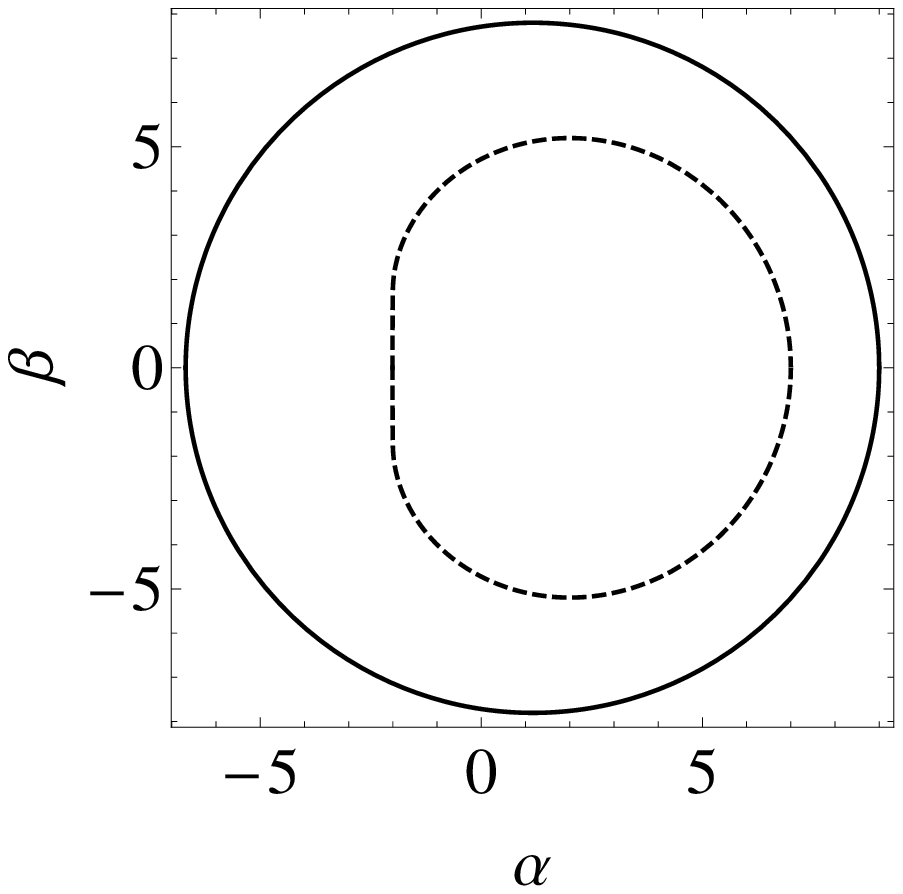} &
            \includegraphics[width=0.25\textwidth]{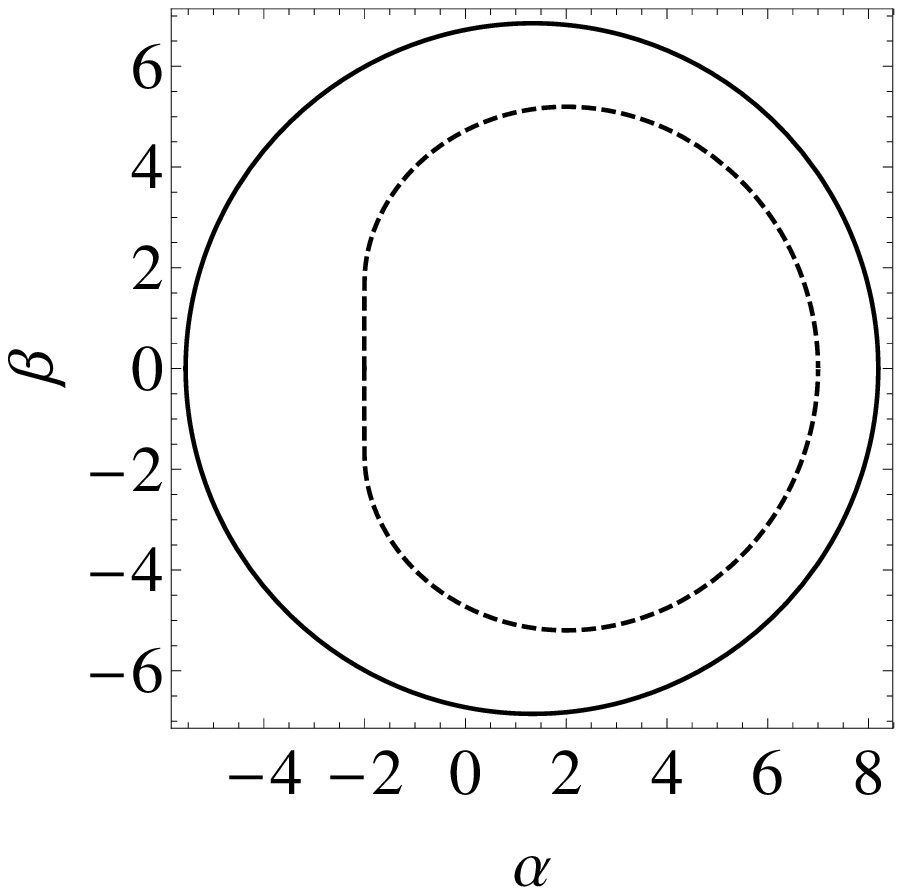} &
            \includegraphics[width=0.25\textwidth]{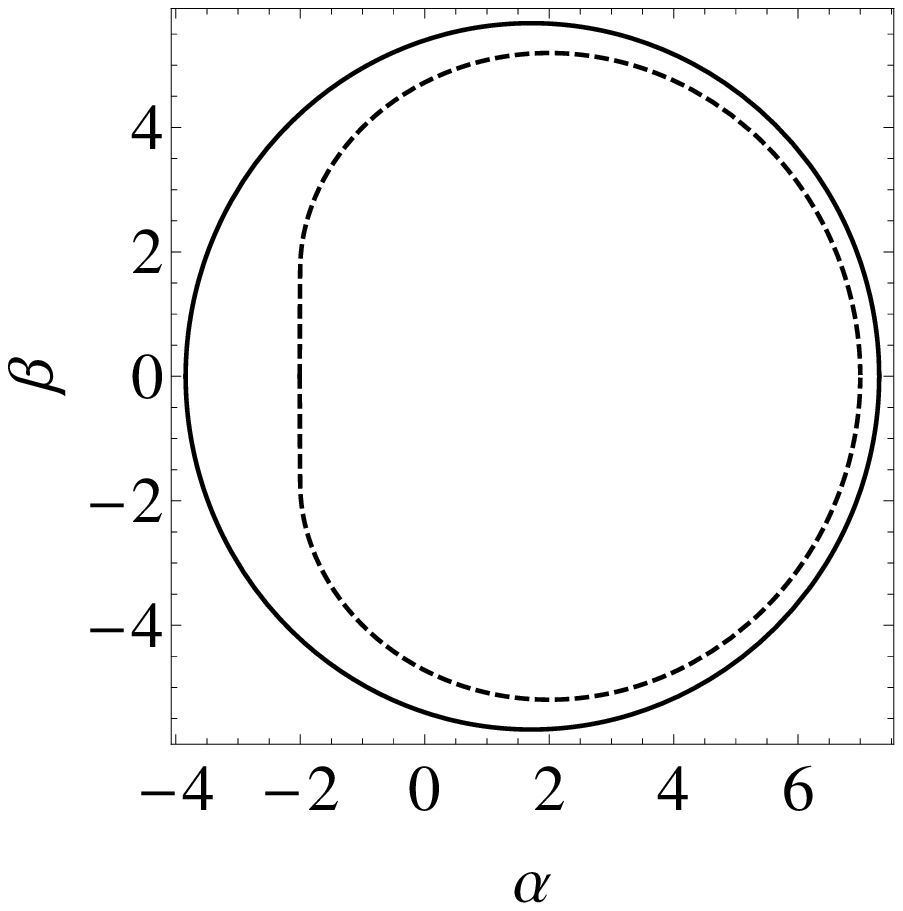} &
            \includegraphics[width=0.25\textwidth]{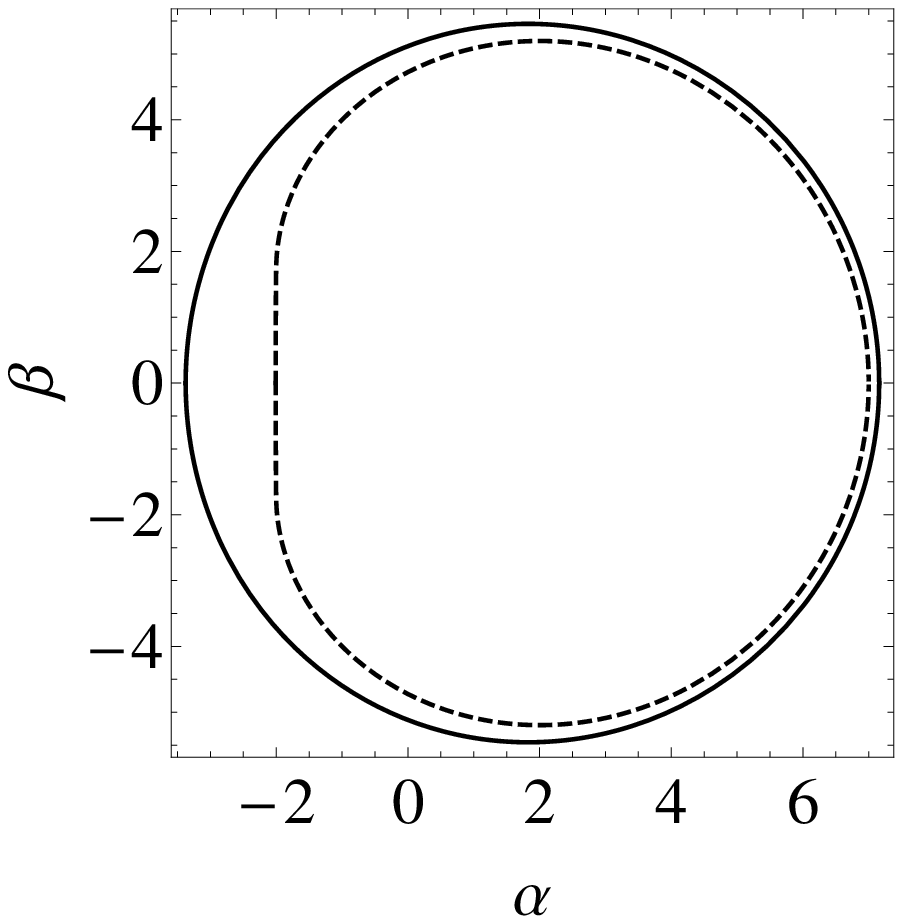} \\
            $a/M=1$, $l/M^{2}=-10$;\  &
            $a/M=1$, $l/M^{2}=-5$;\  &
            $a/M=1$, $l/M^{2}=-1$;\  &
            $a/M=1$, $l/M^{2}=-0.5$ \\
            \includegraphics[width=0.25\textwidth]{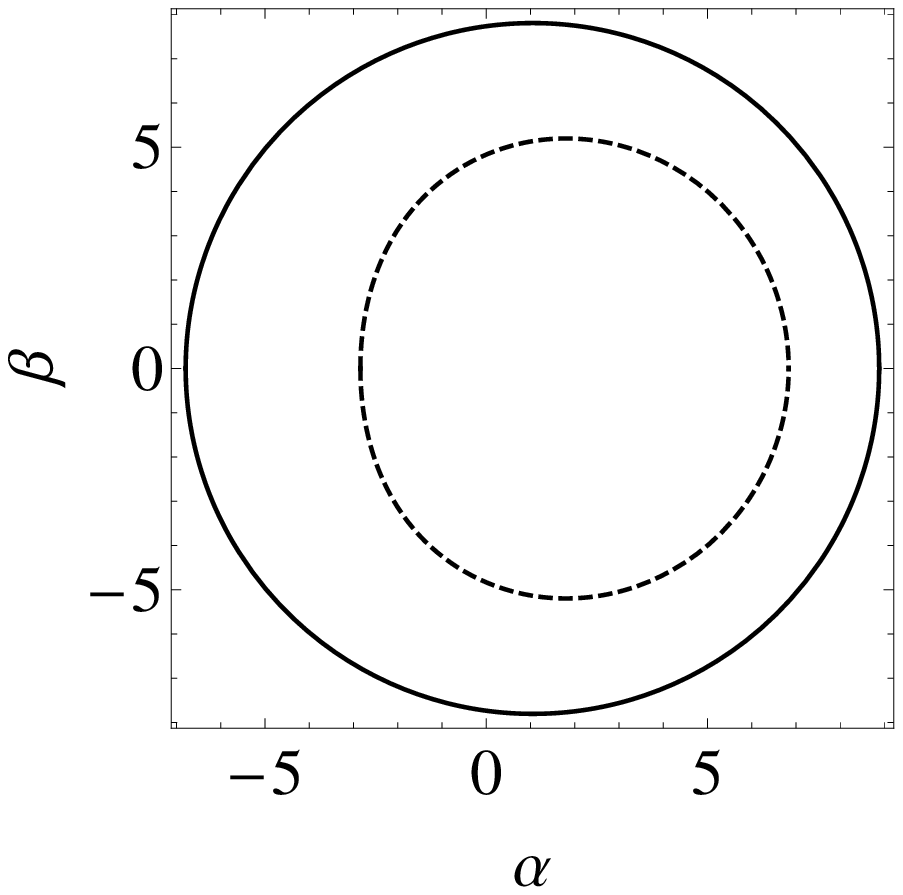} &
            \includegraphics[width=0.25\textwidth]{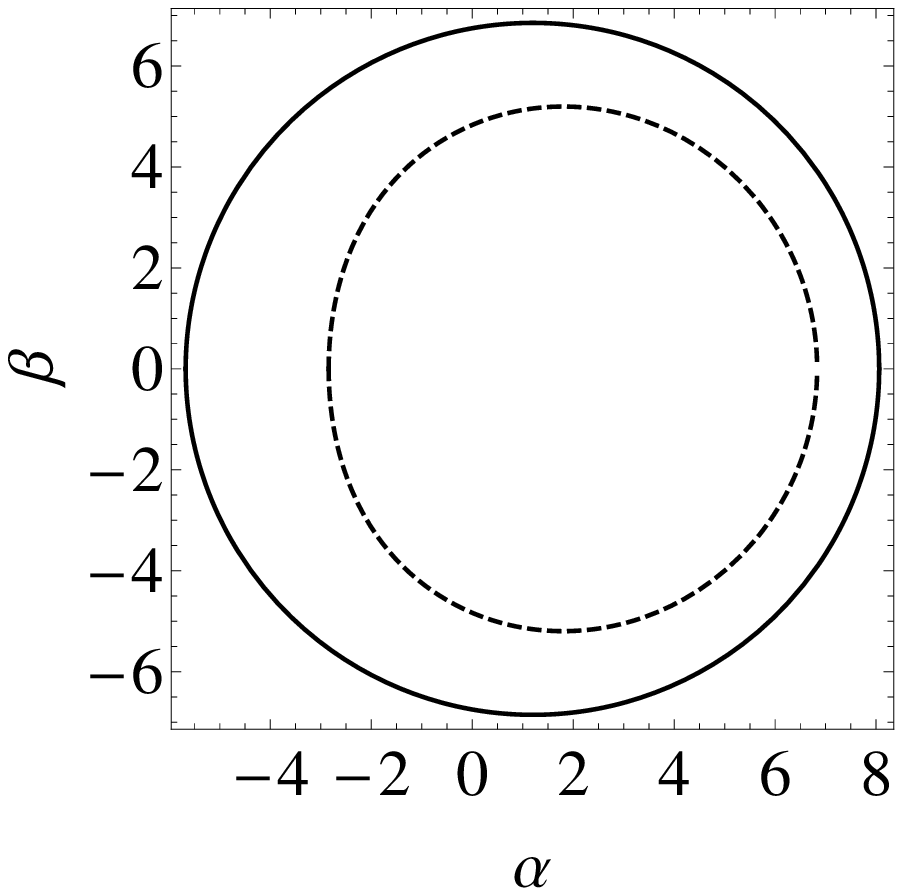} &
            \includegraphics[width=0.25\textwidth]{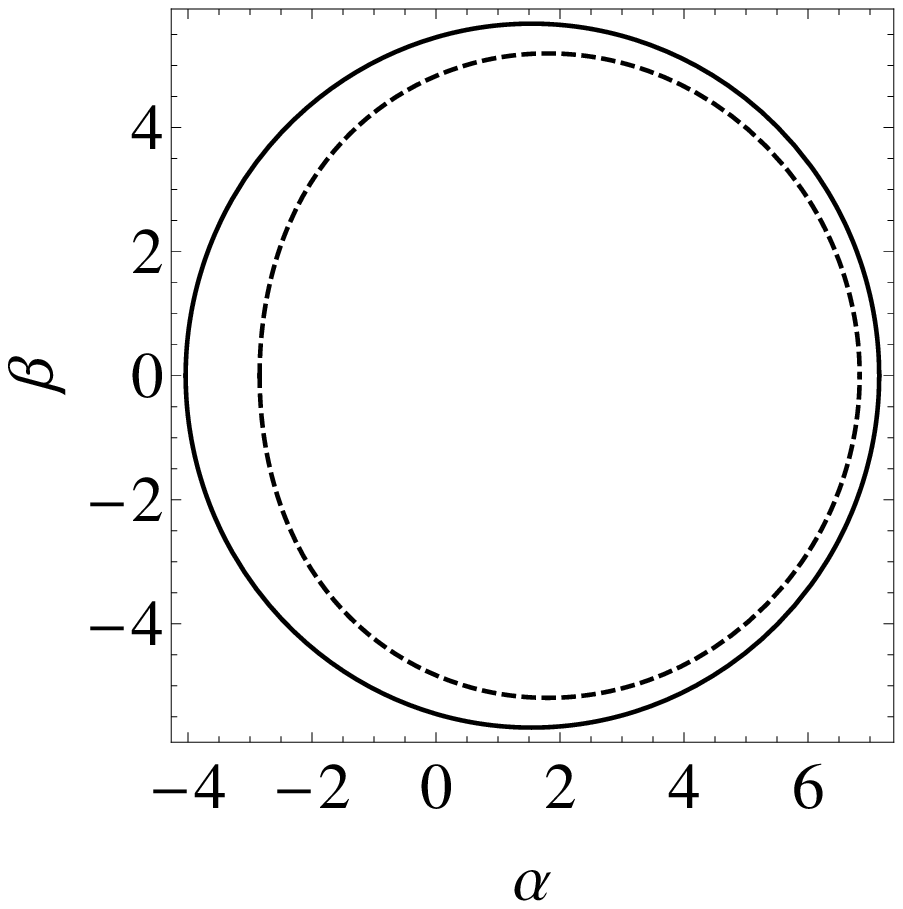} &
            \includegraphics[width=0.25\textwidth]{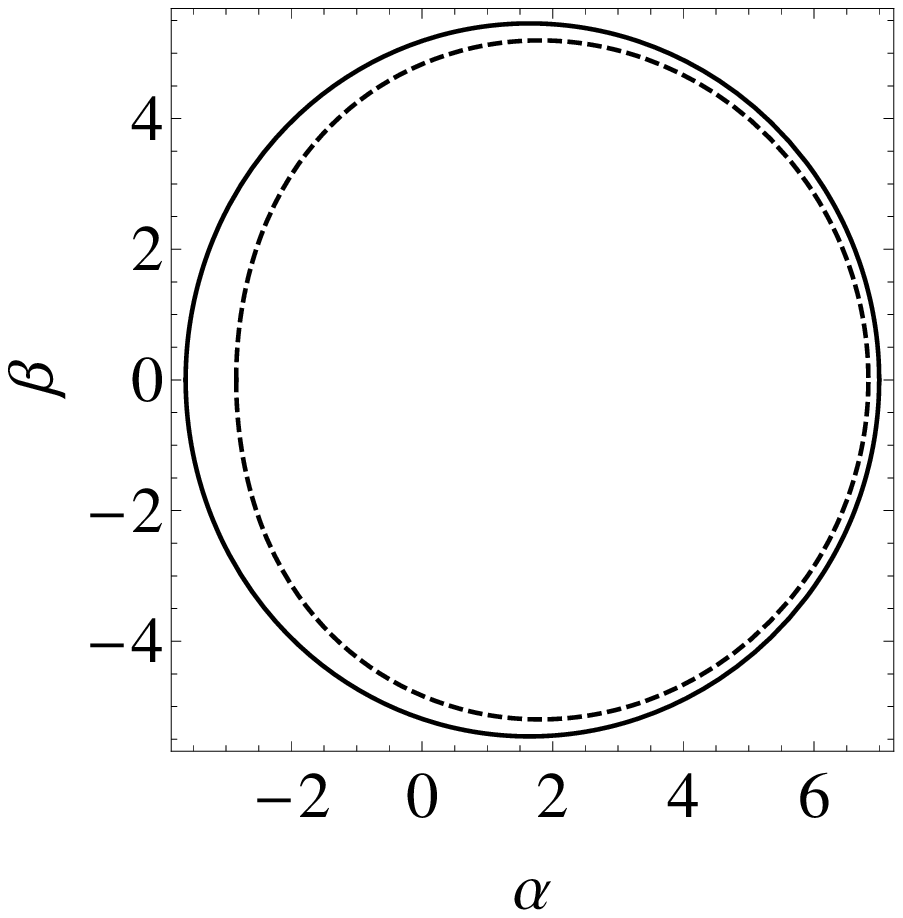} \\
            $a/M=0.9$, $l/M^{2}=-10$;\  &
            $a/M=0.9$, $l/M^{2}=-5$;\  &
            $a/M=0.9$, $l/M^{2}=-1$;\  &
            $a/M=0.9$, $l/M^{2}=-0.5$ \\
            \includegraphics[width=0.25\textwidth]{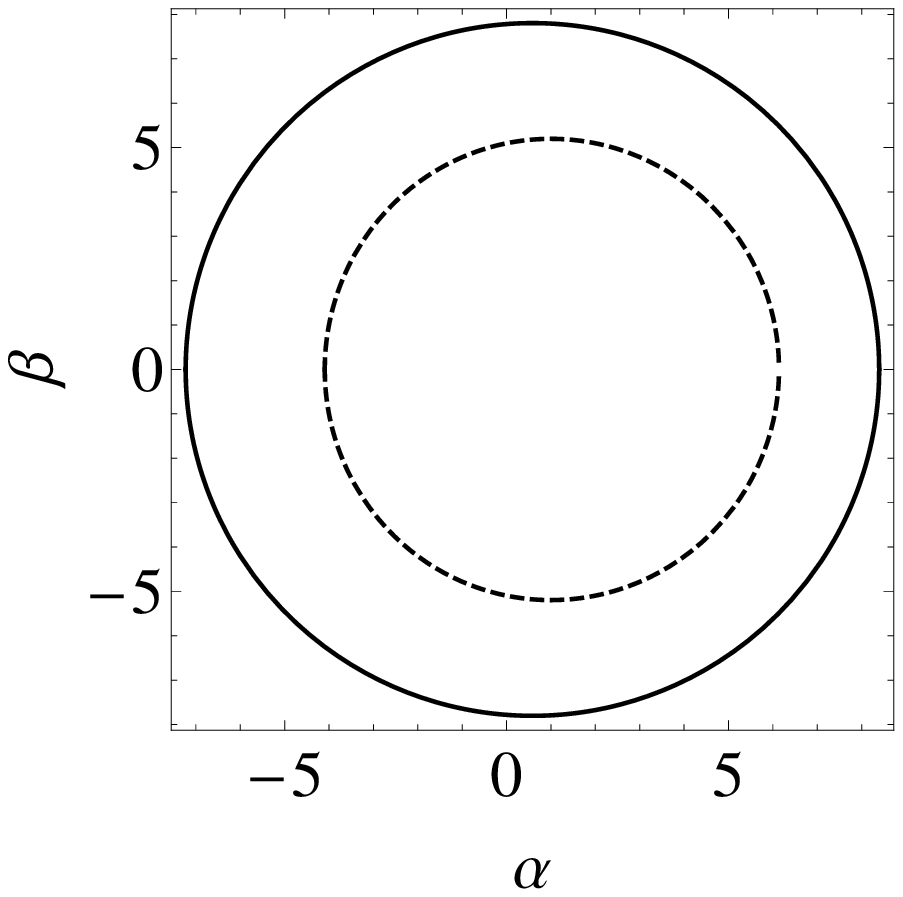} &
            \includegraphics[width=0.25\textwidth]{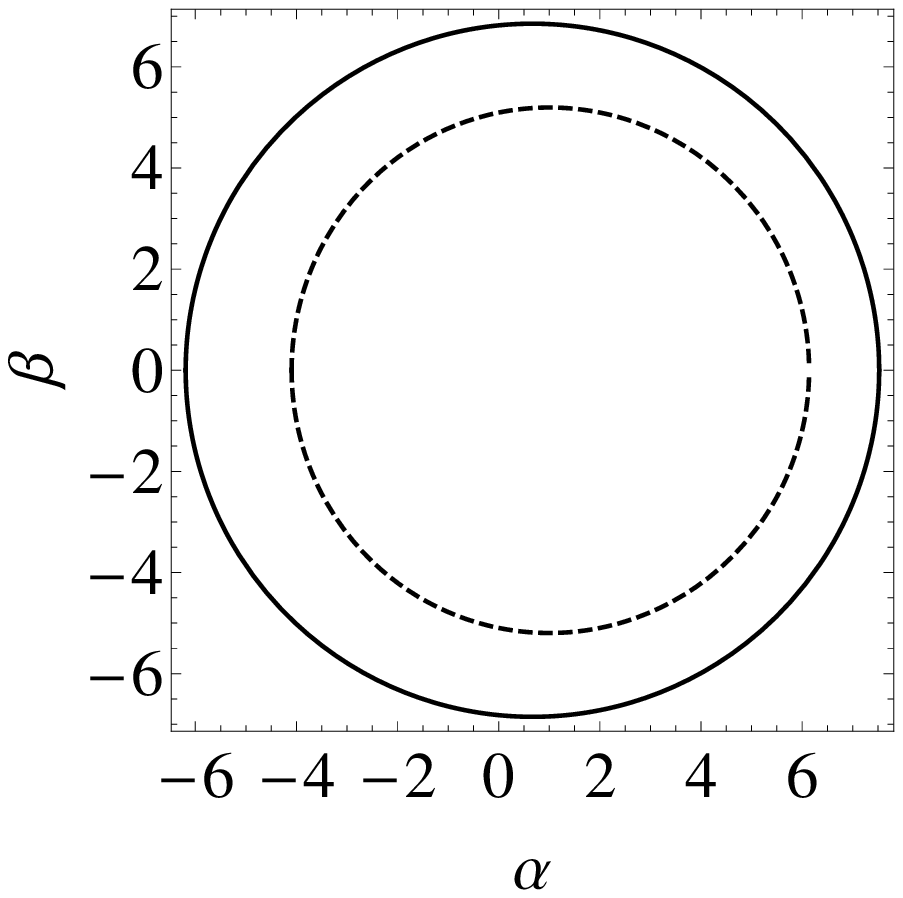} &
            \includegraphics[width=0.25\textwidth]{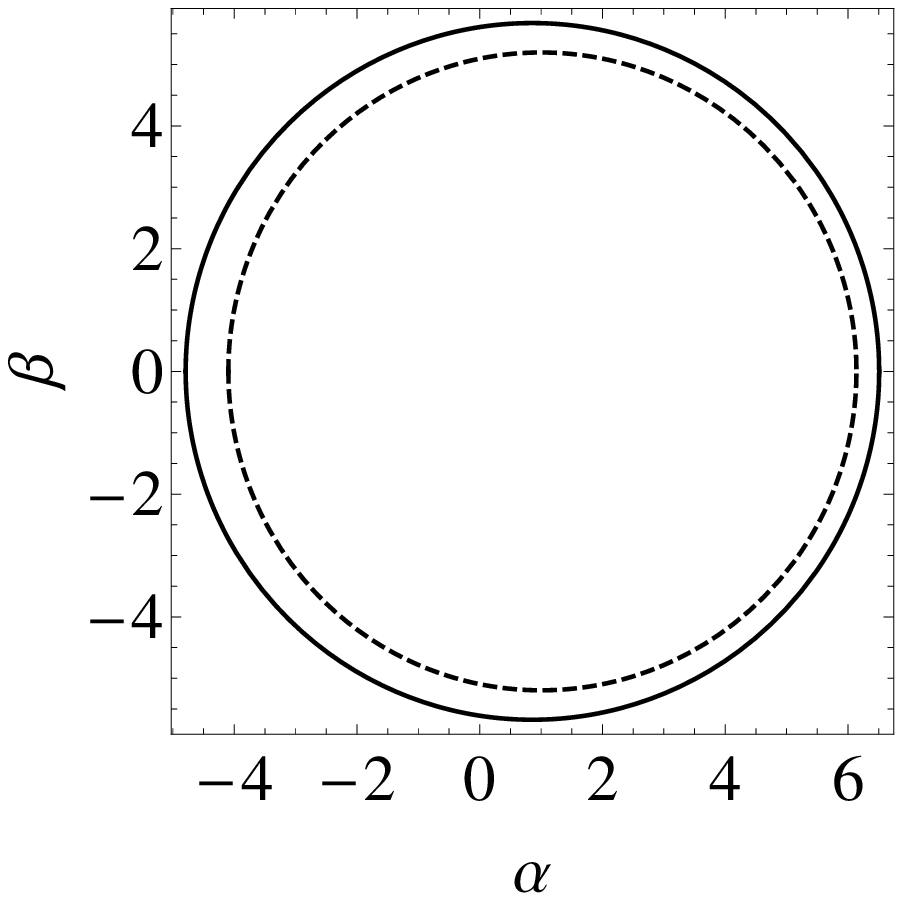} &
            \includegraphics[width=0.25\textwidth]{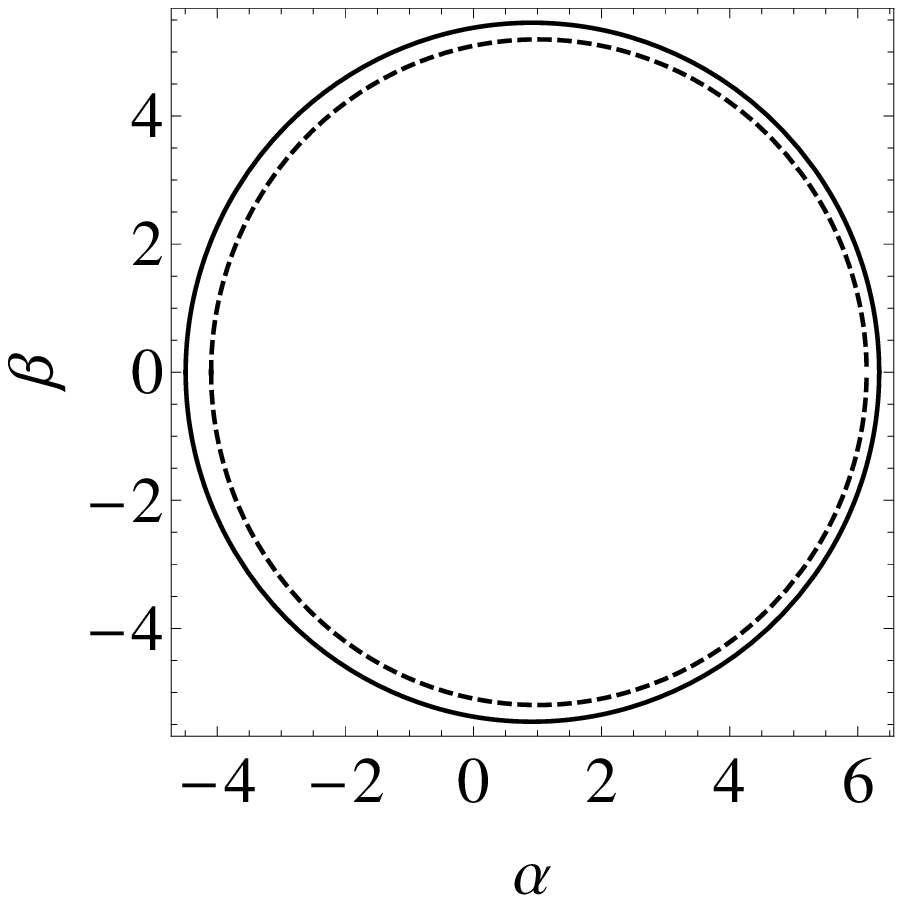} \\
            $a/M=0.5$, $l/M^{2}=-10$;\  &
            $a/M=0.5$, $l/M^{2}=-5$;\  &
            $a/M=0.5$, $l/M^{2}=-1$;\  &
            $a/M=0.5$, $l/M^{2}=-0.5$ \\
            \includegraphics[width=0.25\textwidth]{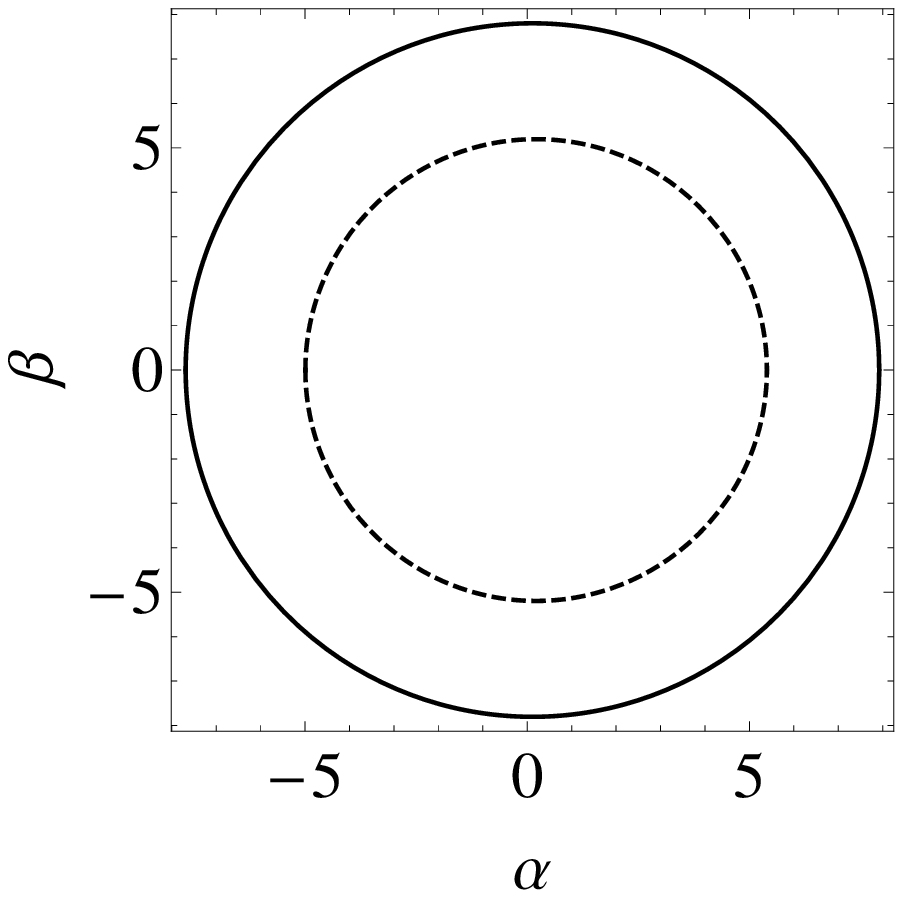} &
            \includegraphics[width=0.25\textwidth]{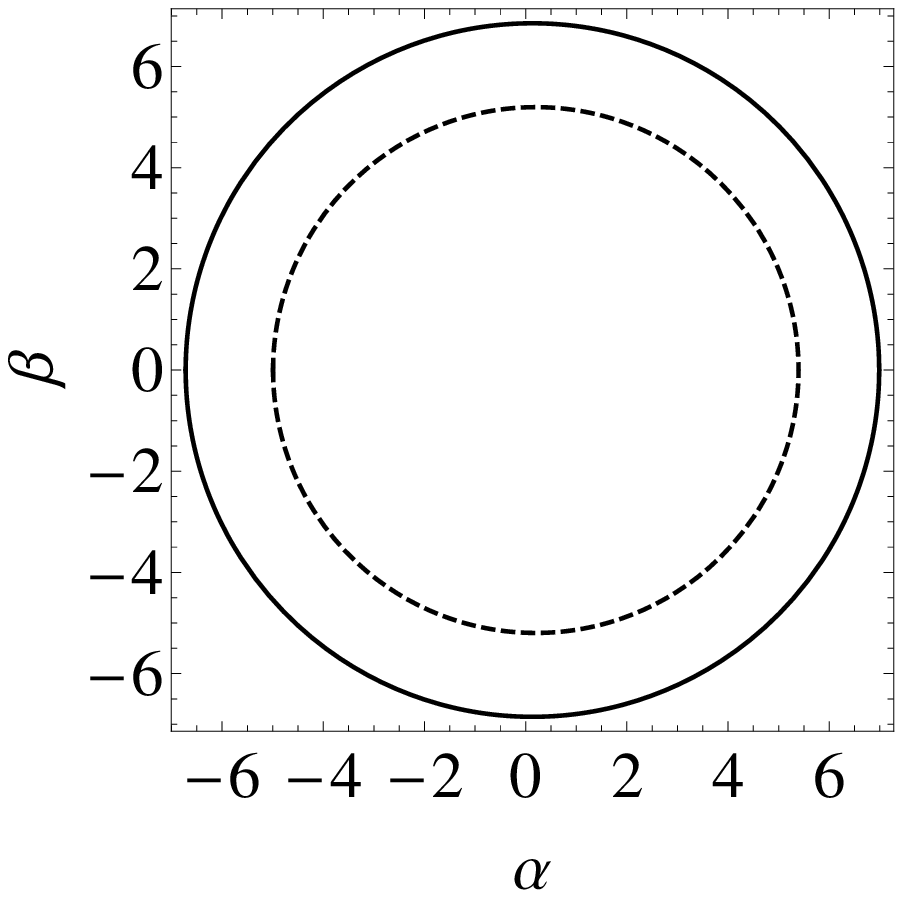} &
            \includegraphics[width=0.25\textwidth]{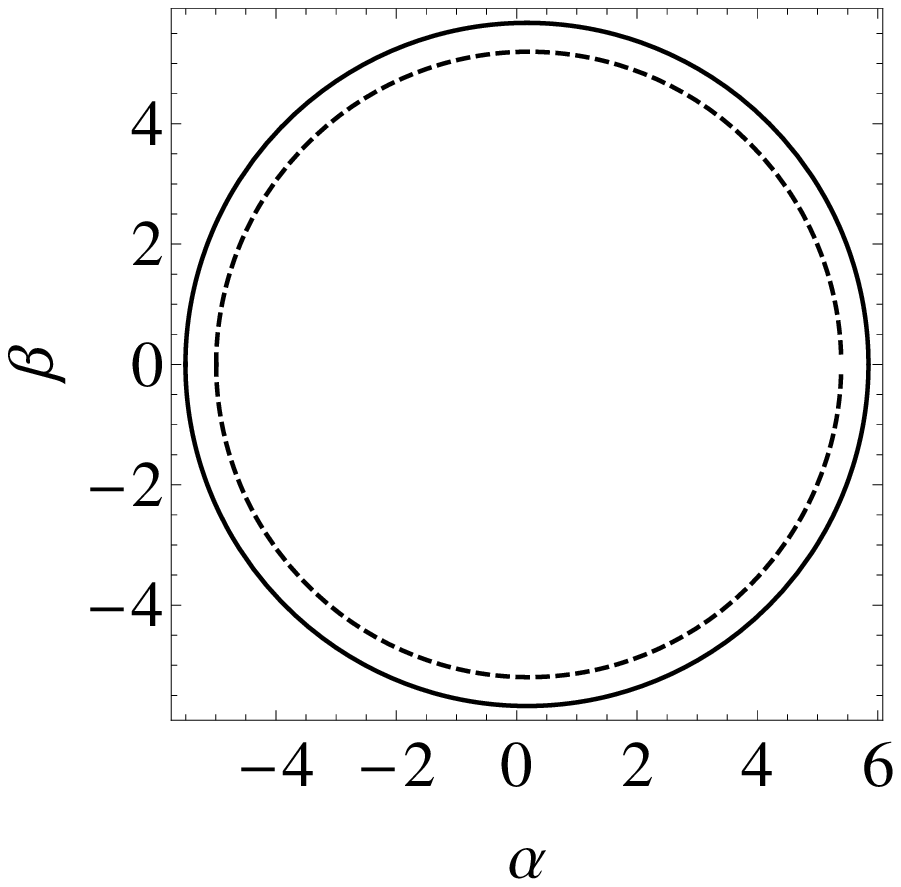} &
            \includegraphics[width=0.25\textwidth]{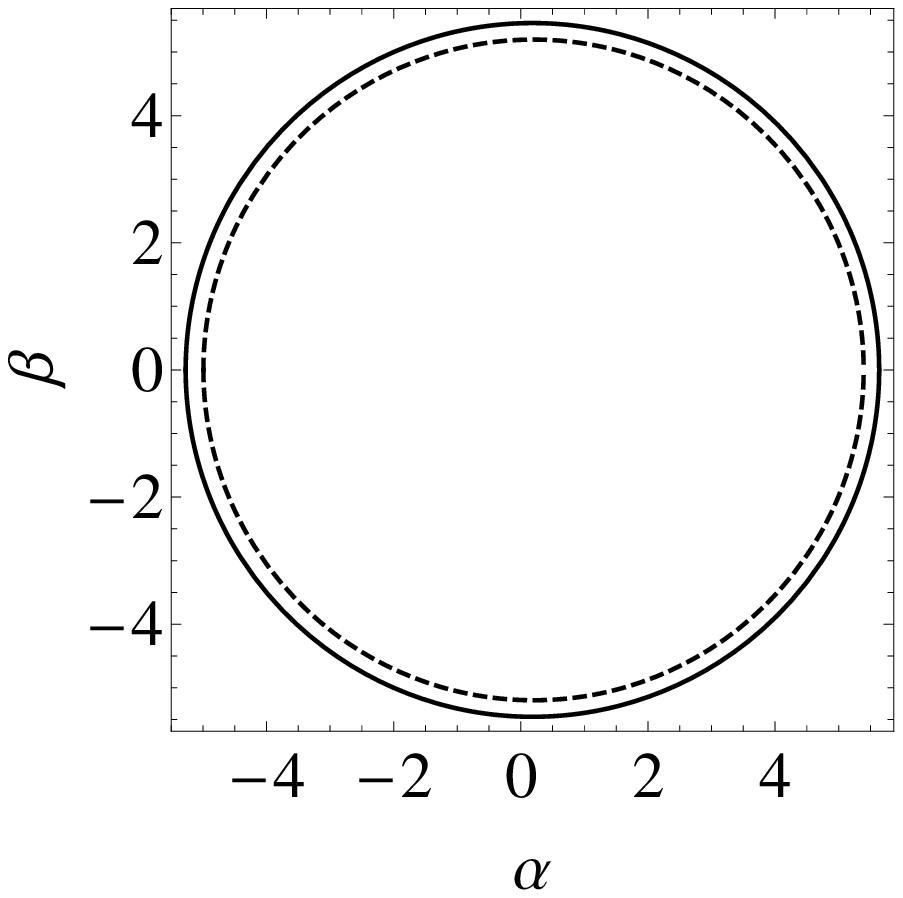} \\
            $a/M=0.1$, $l/M^{2}=-10$;\  &
            $a/M=0.1$, $l/M^{2}=-5$;\  &
            $a/M=0.1$, $l/M^{2}=-1$;\  &
            $a/M=0.1$, $l/M^{2}=-0.5$ \\
        \end{tabular}}
\caption{\footnotesize{The shadow of the generalized Kerr black hole in the model with function $\mu(r)=M e^{-l/r^{2}}$ (solid line) and the Kerr black hole (dashed line)
with inclination angle $\theta_{0}=\pi/2\ rad$ for different
values of parameters $a$ and $l$. The parameter $M$ of both solutions
is set equal to 1. The celestial coordinates $(\alpha,\beta)$ are
measured in the units of parameter $M$. } }
        \label{WS_a3minus}
\end{figure}

\begin{figure}[h]
        \setlength{\tabcolsep}{ 0 pt }{\scriptsize\tt
        \begin{tabular}{ cccc }
            \includegraphics[width=0.25\textwidth]{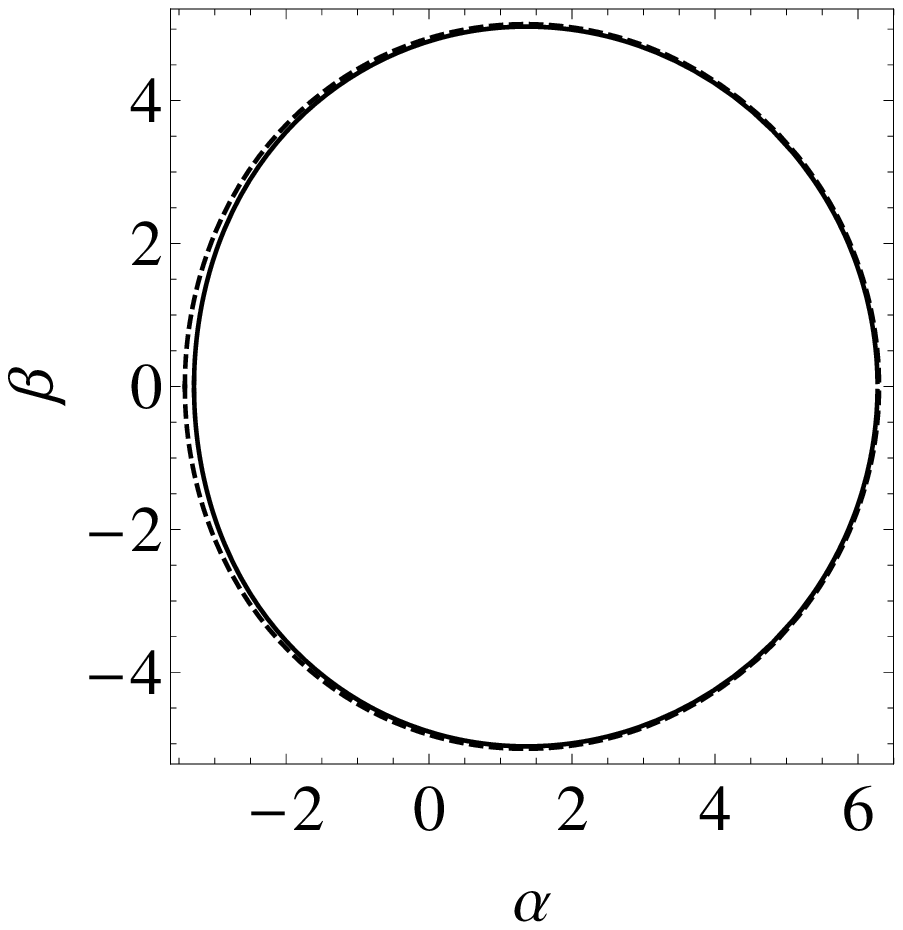} &
            \includegraphics[width=0.25\textwidth]{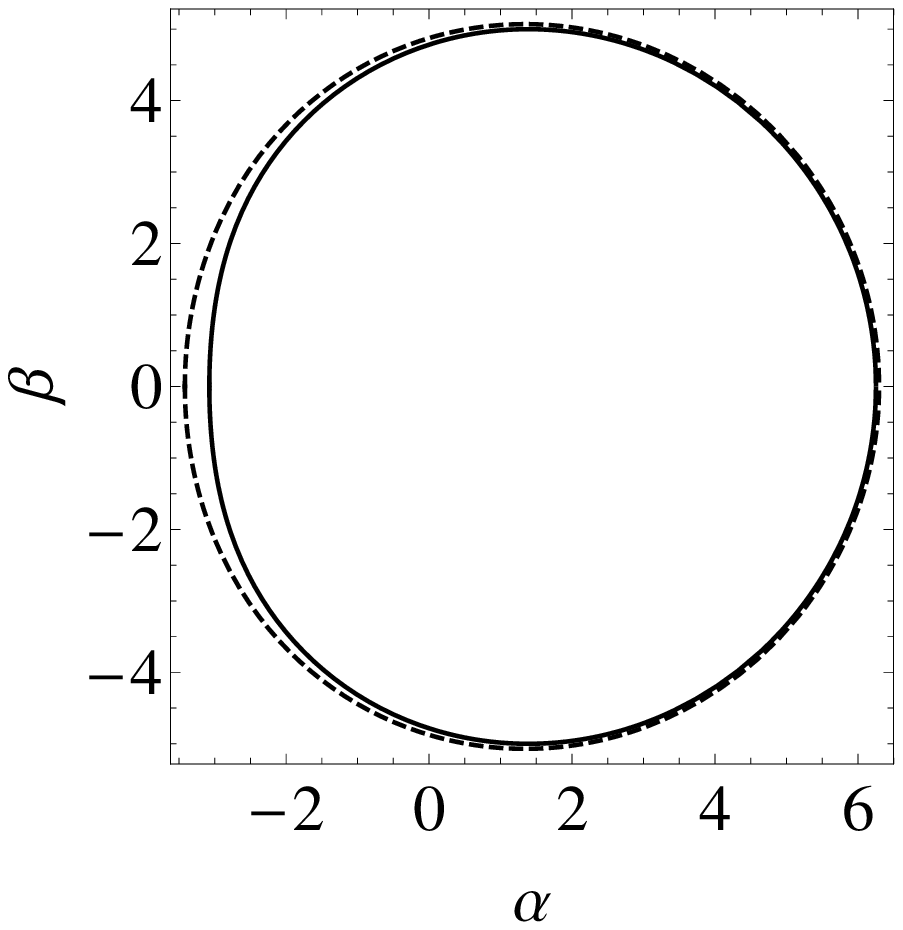} &
            \includegraphics[width=0.25\textwidth]{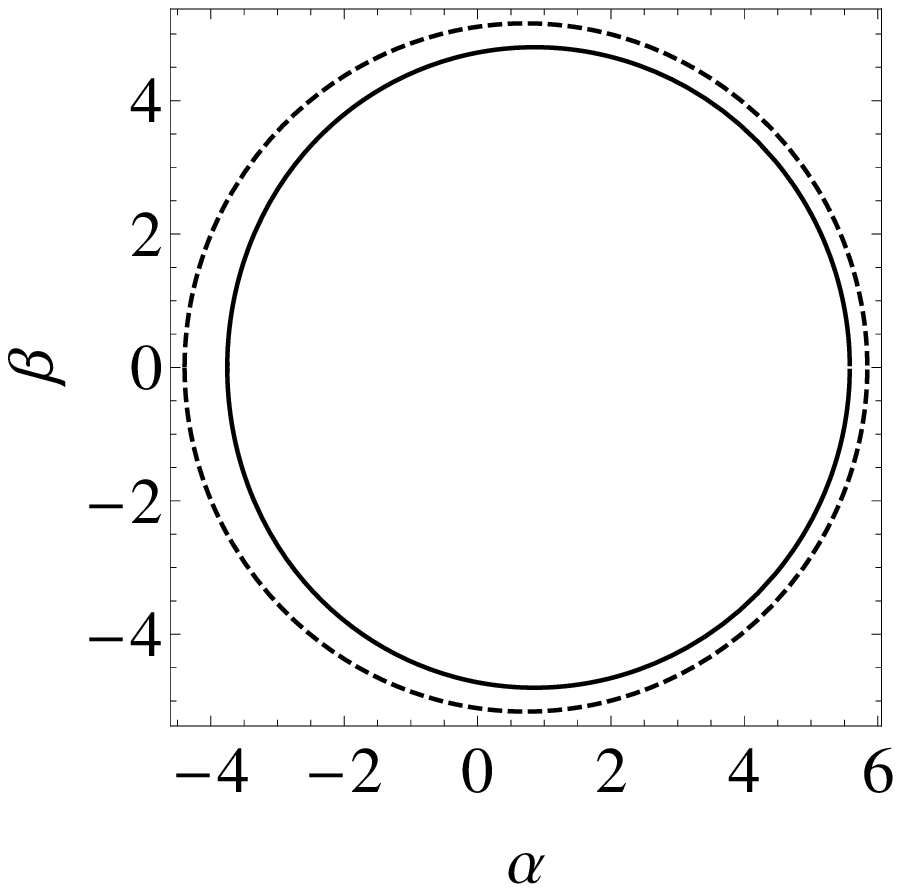} &
            \includegraphics[width=0.25\textwidth]{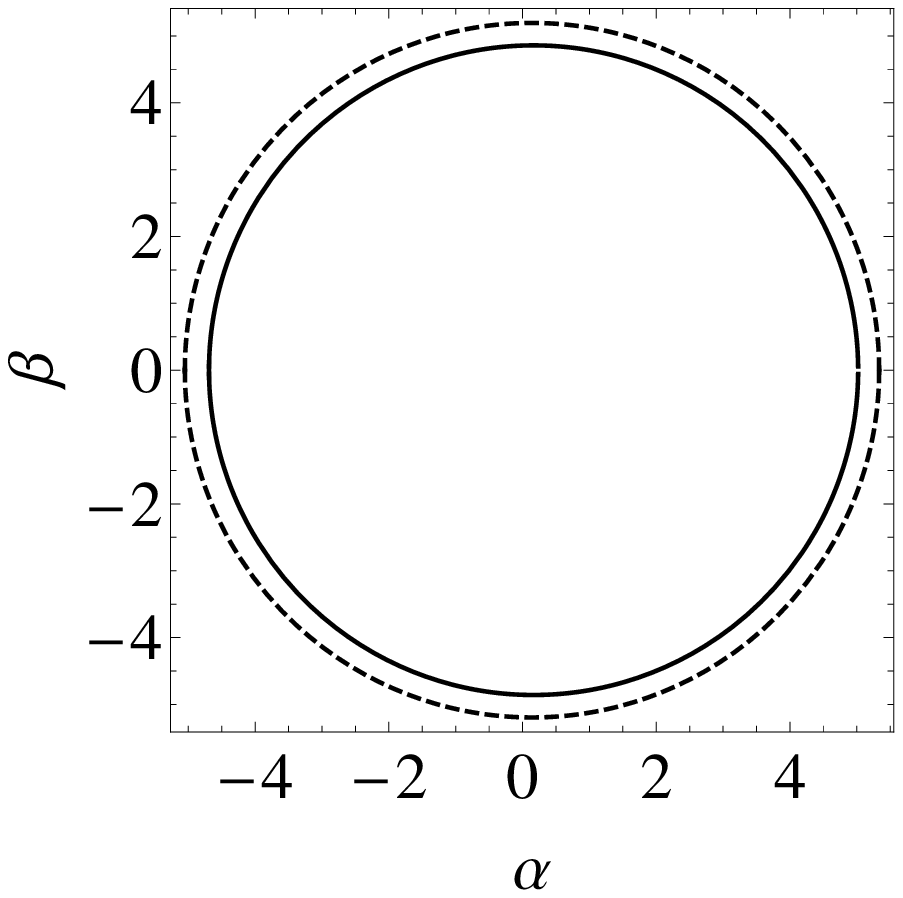} \\
            $a/M=0.9$, $l/M^{2}=0.05$;\  &
            $a/M=0.9$, $l/M^{2}=0.1$;\  &
            $a/M=0.5$, $l/M^{2}=0.5$;\  &
            $a/M=0.1$, $l/M^{2}=0.5$ \\
        \end{tabular}}
\caption{\footnotesize{The shadow of the generalized Kerr black hole in the model with function $\mu(r)=M e^{-l/r^{2}}$ (solid line) and the Kerr black hole (dashed line)
with inclination angle $\theta_{0}=\pi/4\ rad$ for different
values of parameters $a$ and $l$. The parameter $M$ of both solutions
is set equal to 1. The celestial coordinates $(\alpha,\beta)$ are
measured in the units of parameter $M$. } }
        \label{WS_a4}
\end{figure}

\begin{figure}[h]
        \setlength{\tabcolsep}{ 0 pt }{\scriptsize\tt
        \begin{tabular}{ cccc }
            \includegraphics[width=0.25\textwidth]{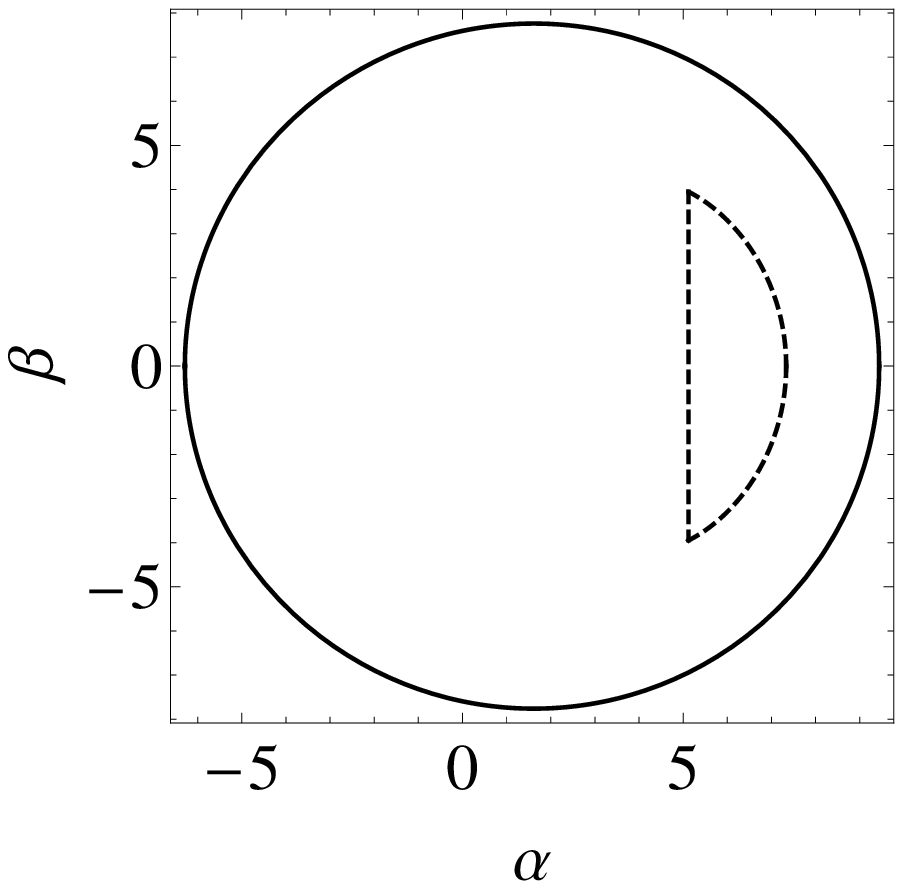} &
            \includegraphics[width=0.25\textwidth]{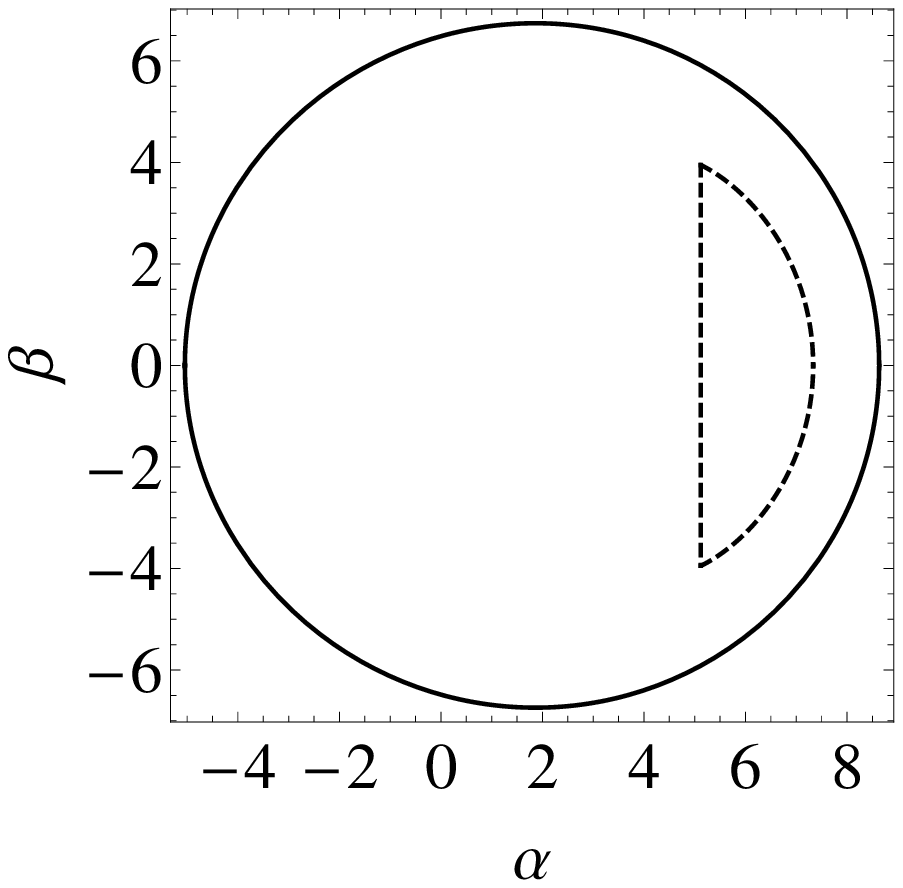} &
            \includegraphics[width=0.25\textwidth]{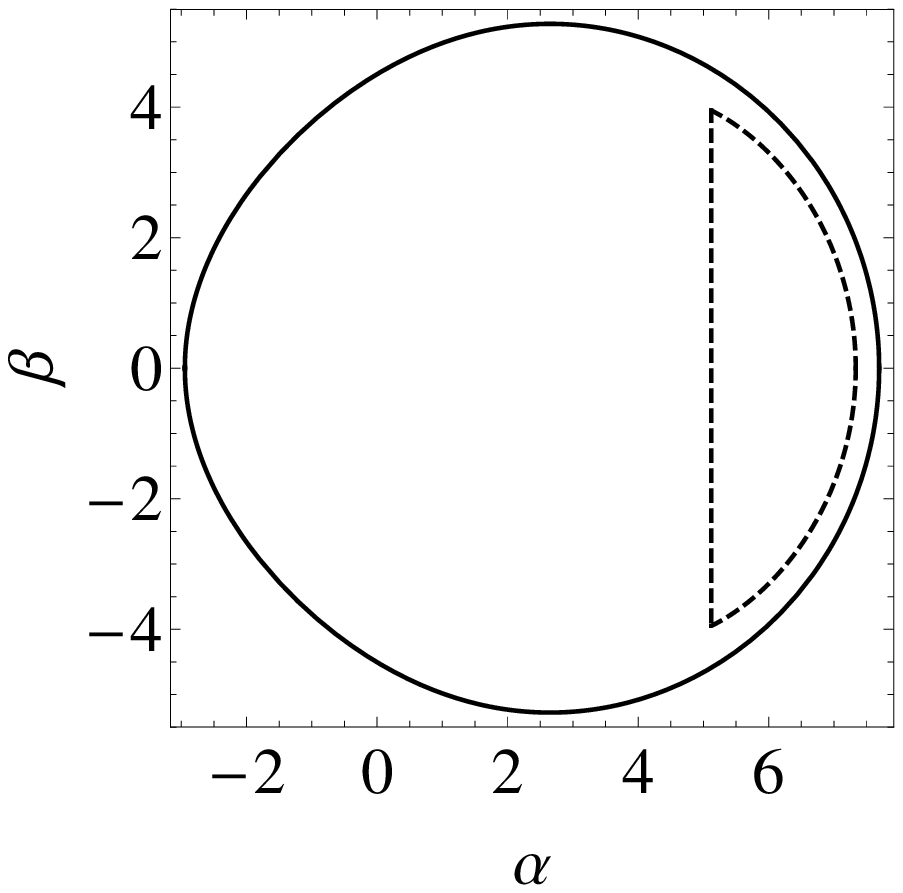} &
            \includegraphics[width=0.25\textwidth]{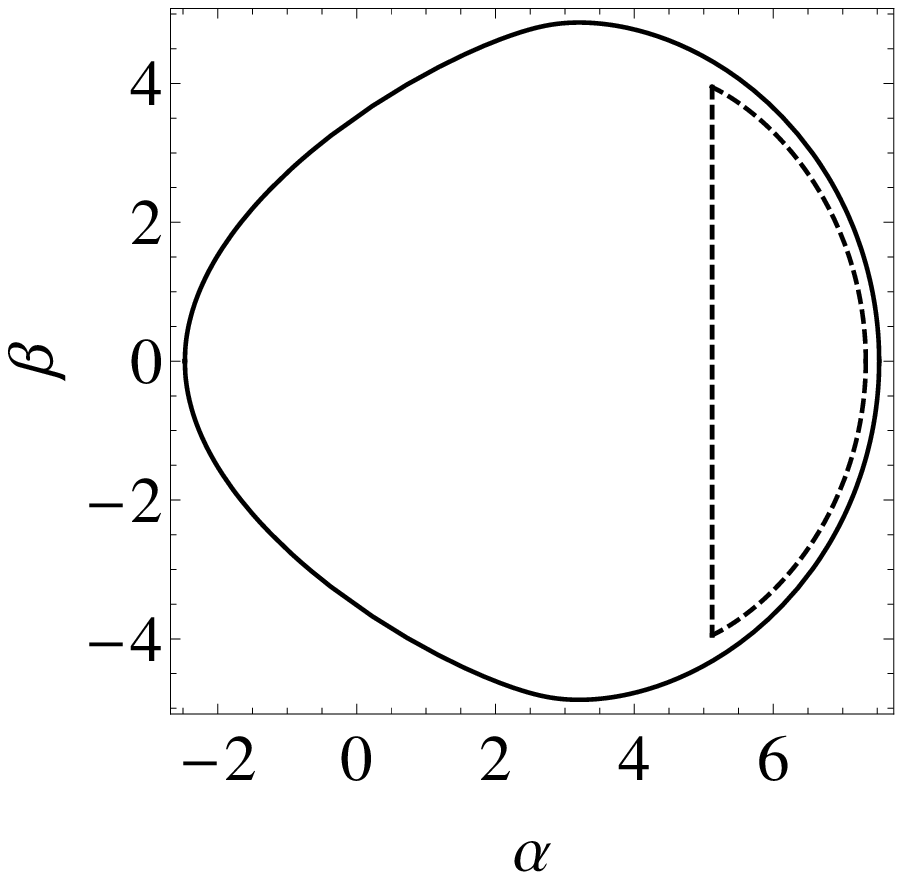} \\
            $a/M=2$, $l/M^{2}=-10$;\  &
            $a/M=2$, $l/M^{2}=-5$;\  &
            $a/M=2$, $l/M^{2}=-1$;\  &
            $a/M=2$, $l/M^{2}=-0.5$ \\
            \includegraphics[width=0.25\textwidth]{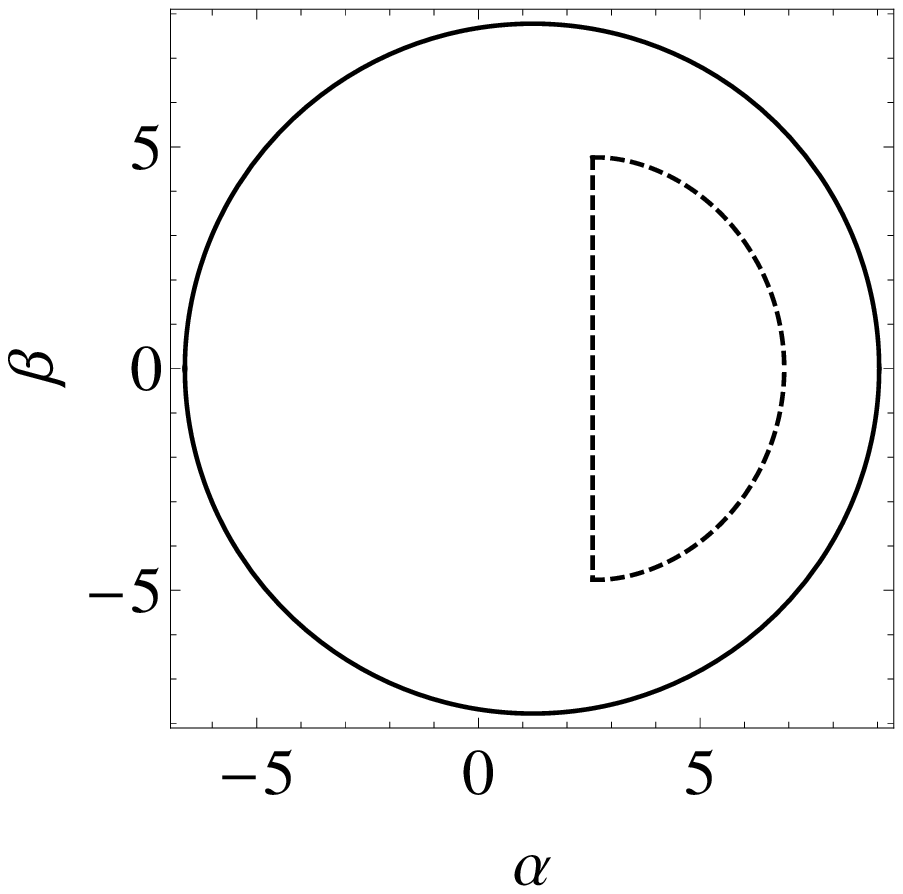} &
            \includegraphics[width=0.25\textwidth]{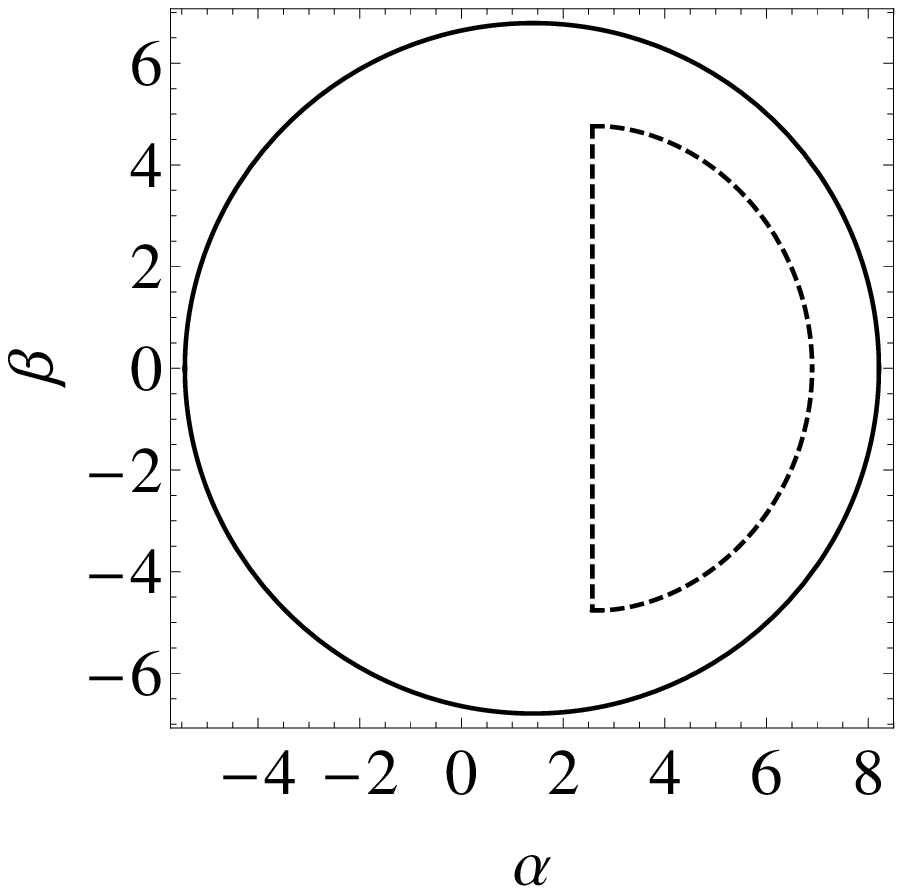} &
            \includegraphics[width=0.25\textwidth]{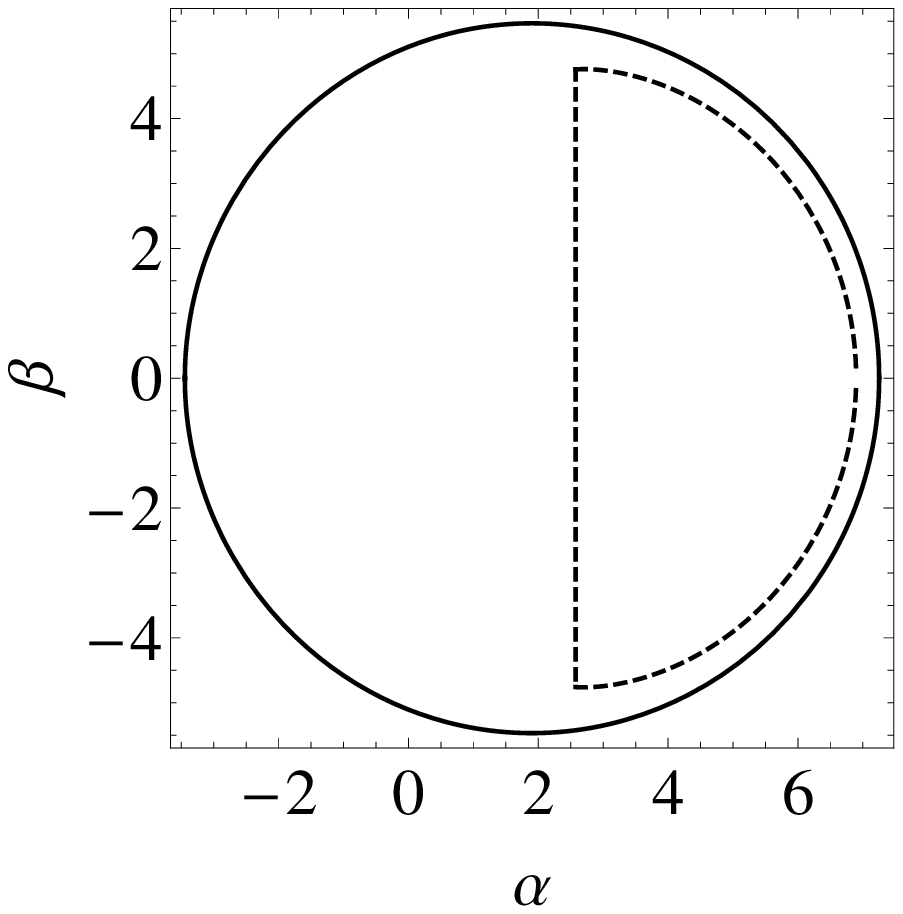} &
            \includegraphics[width=0.25\textwidth]{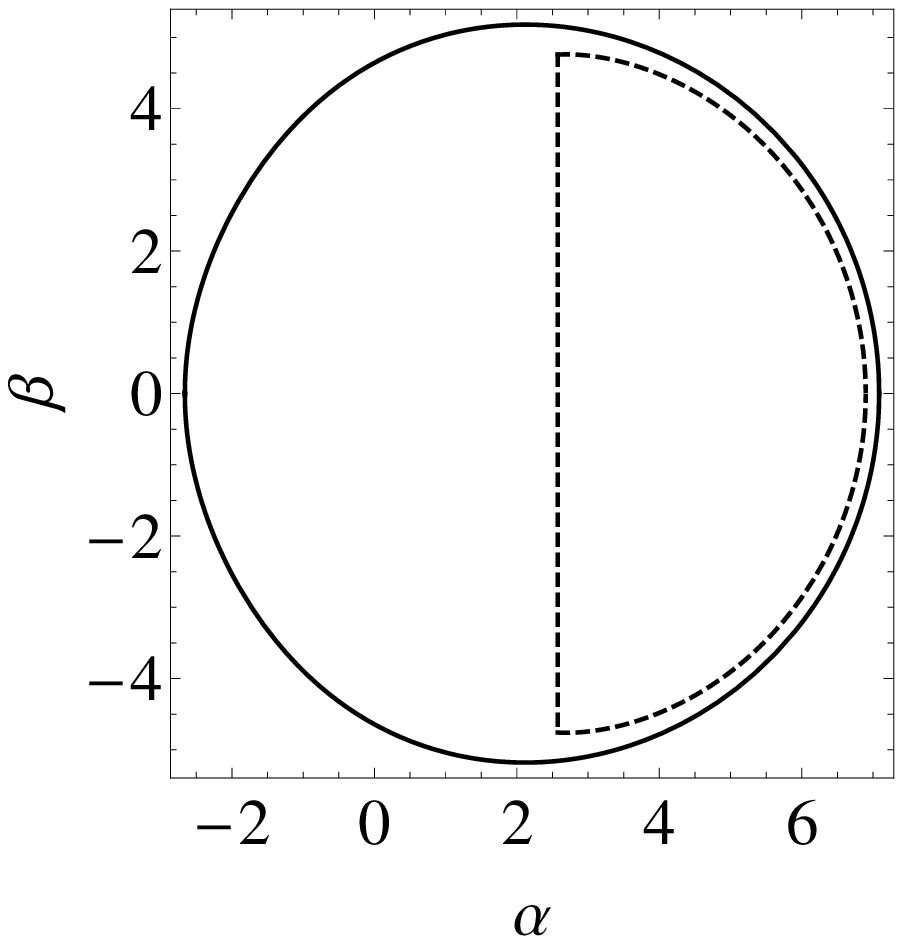} \\
            $a/M=1.5$, $l/M^{2}=-10$;\  &
            $a/M=1.5$, $l/M^{2}=-5$;\  &
            $a/M=1.5$, $l/M^{2}=-1$;\  &
            $a/M=1.5$, $l/M^{2}=-0.5$ \\
        \end{tabular}}
\caption{\footnotesize{The shadow of the generalized Kerr black hole in the model with function $\mu(r)=M e^{-l/r^{2}}$ (solid line) and the corresponding  Kerr naked singularity (dashed line)
with inclination angle $\theta_{0}=\pi/4\ rad$ for different
values of parameters $a$ and $l$. The parameter $M$ of both solutions
is set equal to 1. The celestial coordinates $(\alpha,\beta)$ are
measured in the units of parameter $M$. } }
        \label{WS_a6minus}
\end{figure}

\begin{figure}[h]
        \setlength{\tabcolsep}{ 0 pt }{\scriptsize\tt
        \begin{tabular}{ cccc }
            \includegraphics[width=0.25\textwidth]{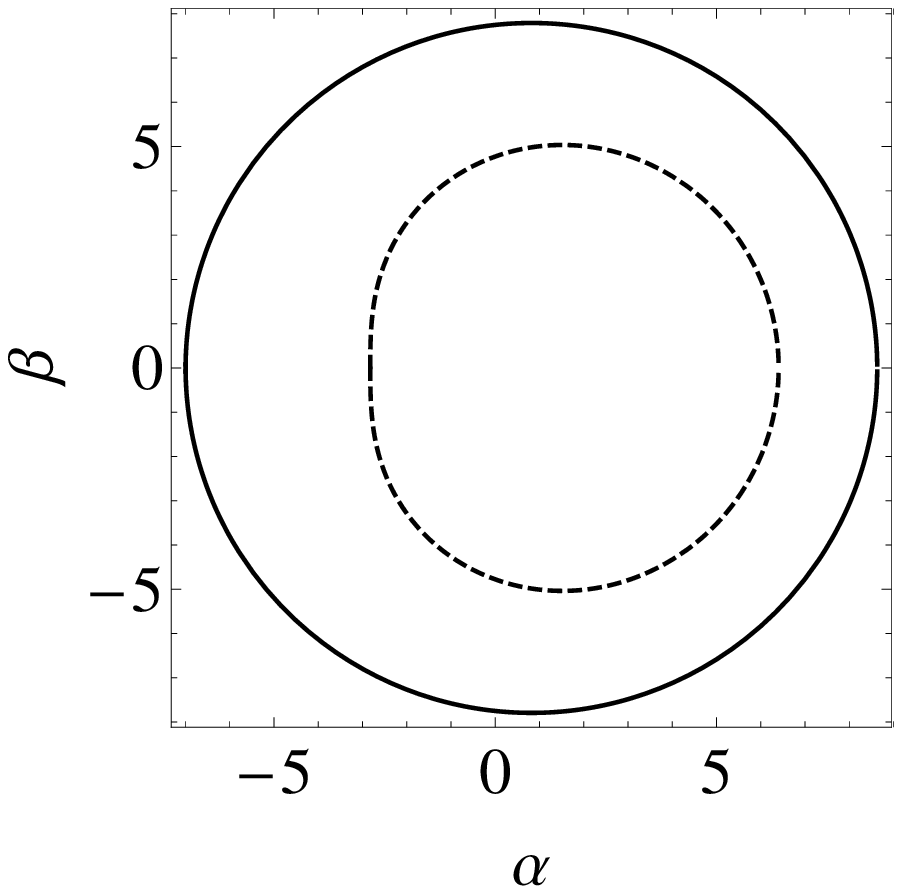} &
            \includegraphics[width=0.25\textwidth]{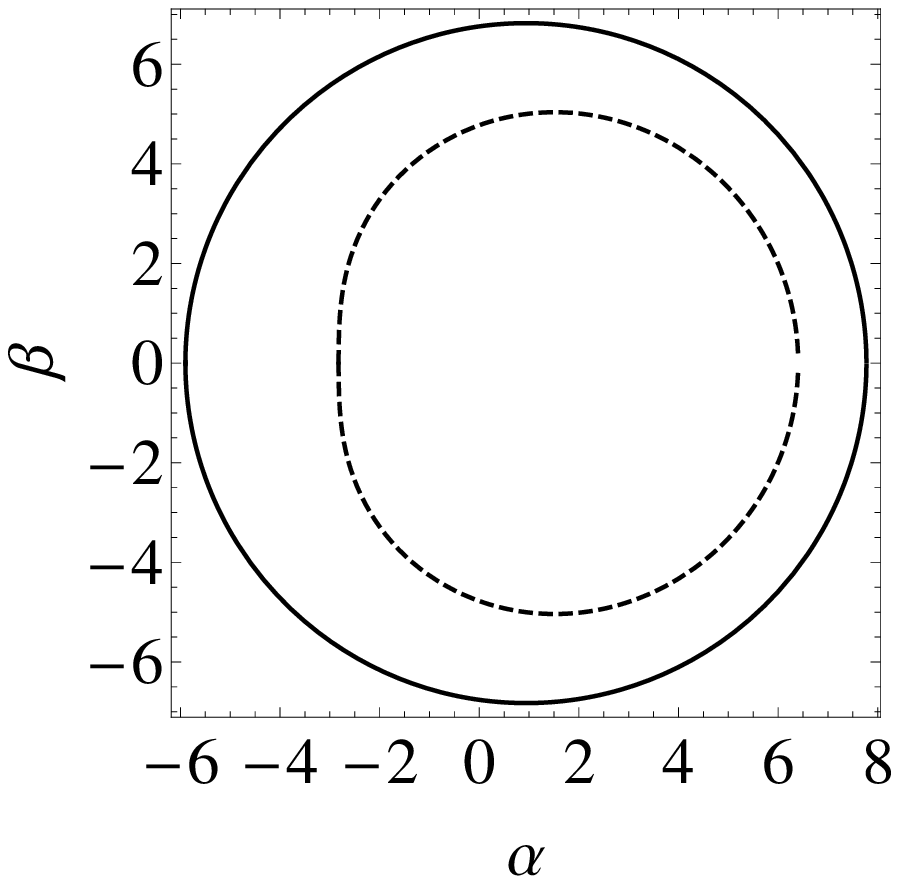} &
            \includegraphics[width=0.25\textwidth]{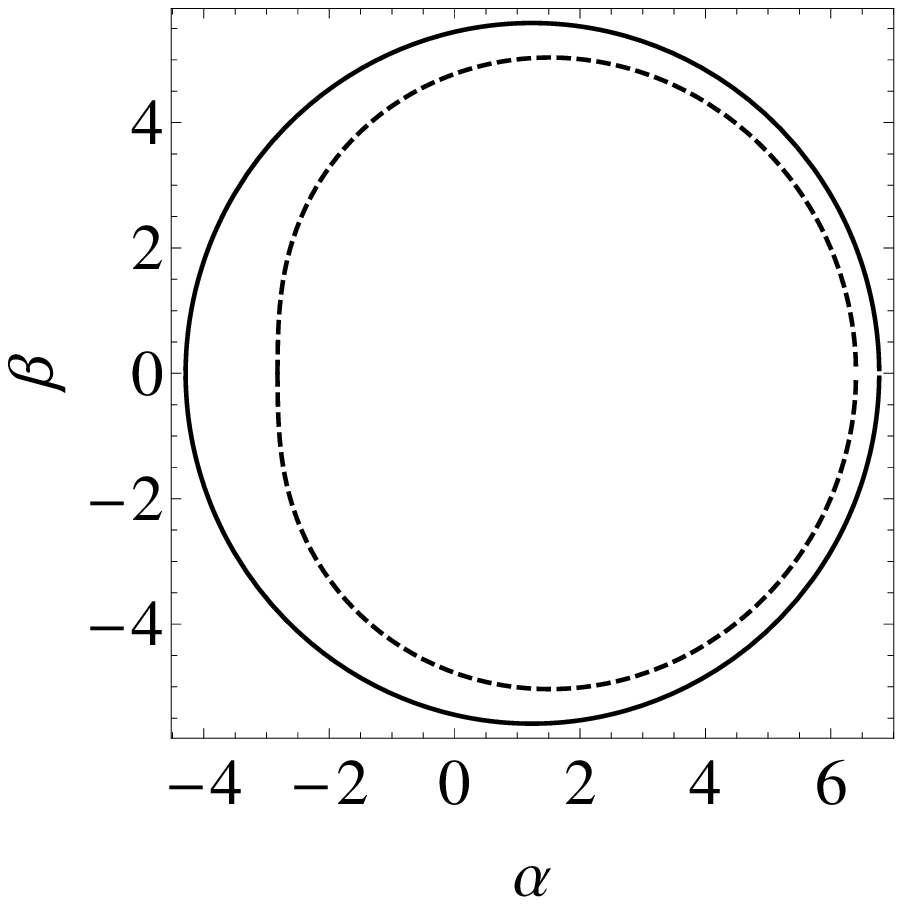} &
            \includegraphics[width=0.25\textwidth]{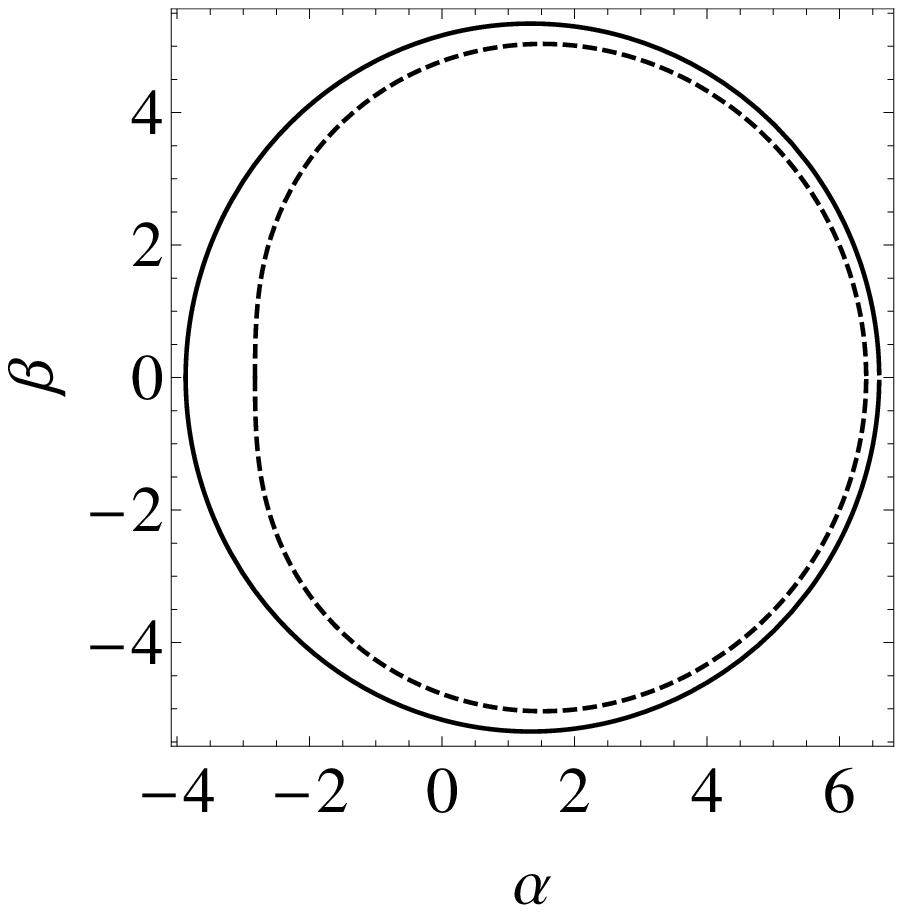} \\
            $a/M=1$, $l/M^{2}=-10$;\  &
            $a/M=1$, $l/M^{2}=-5$;\  &
            $a/M=1$, $l/M^{2}=-1$;\  &
            $a/M=1$, $l/M^{2}=-0.5$ \\
            \includegraphics[width=0.25\textwidth]{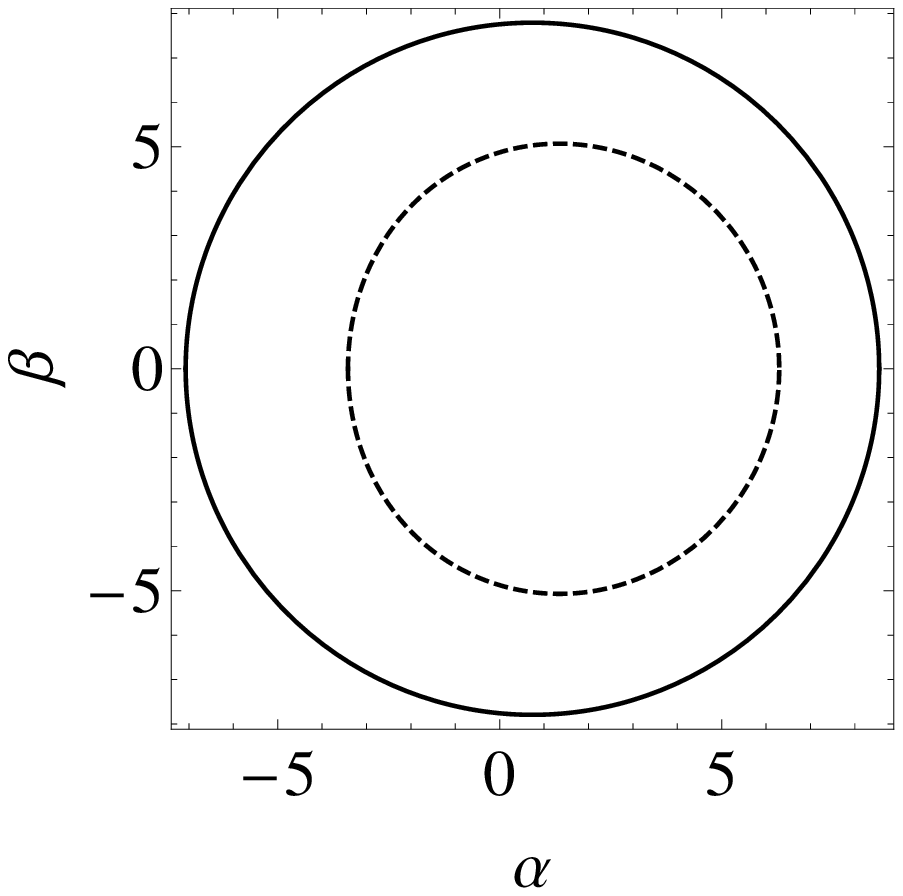} &
            \includegraphics[width=0.25\textwidth]{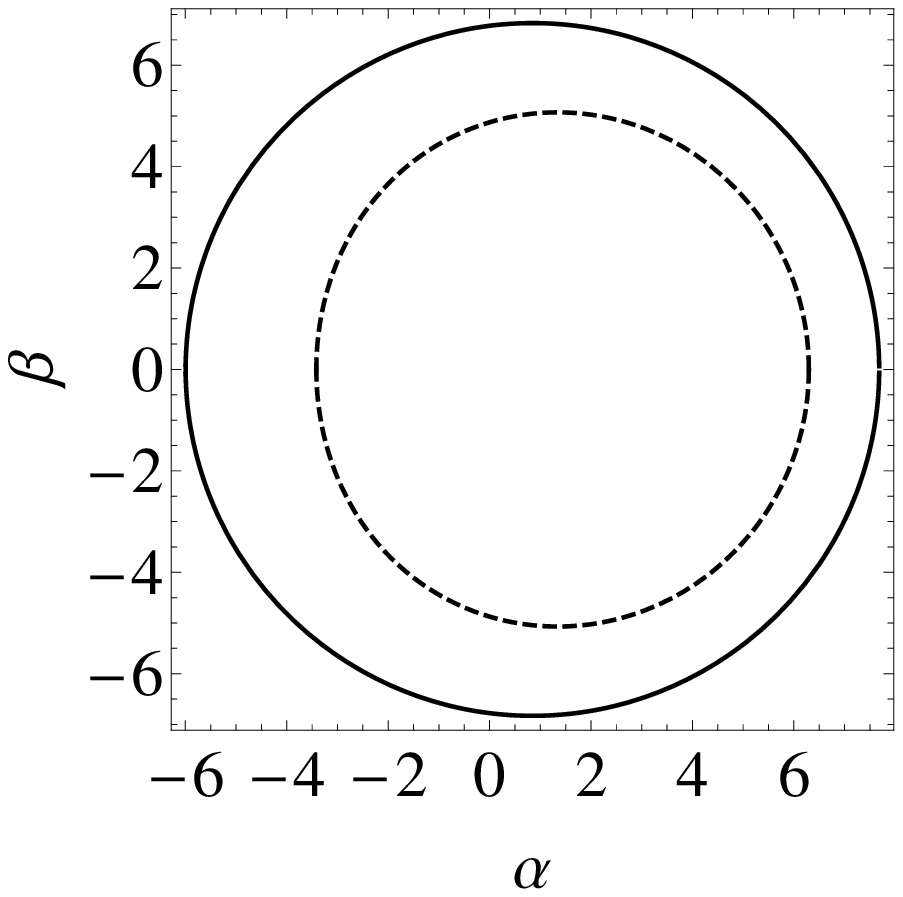} &
            \includegraphics[width=0.25\textwidth]{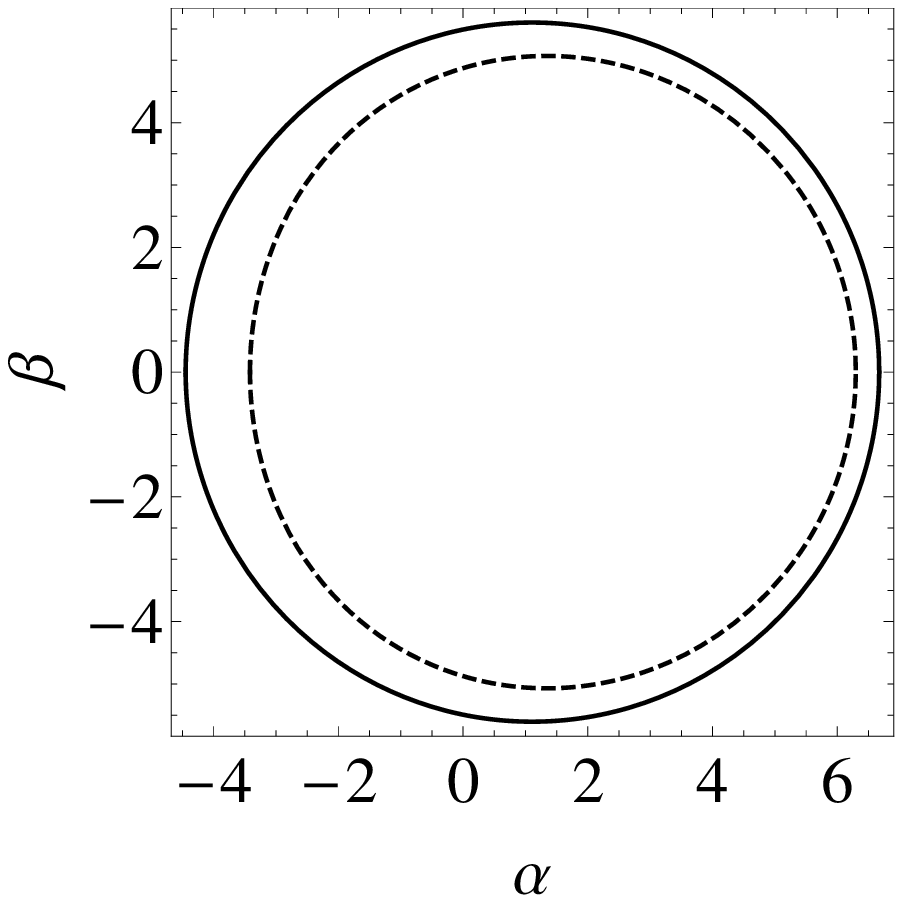} &
            \includegraphics[width=0.25\textwidth]{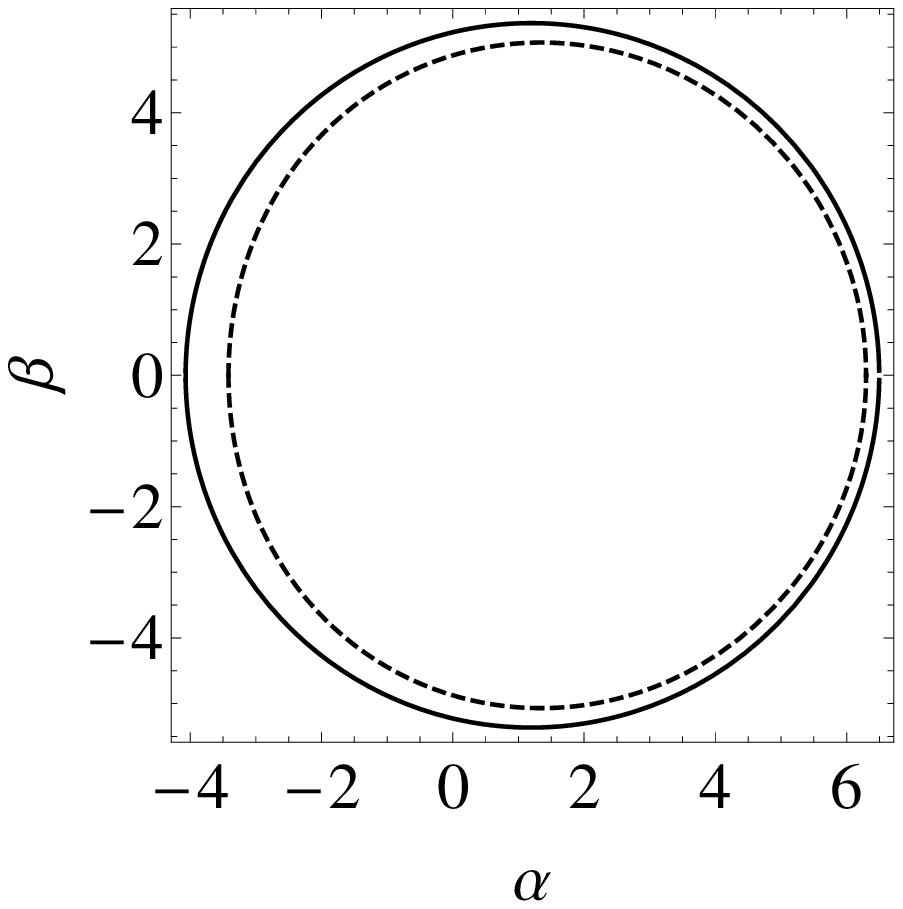} \\
            $a/M=0.9$, $l/M^{2}=-10$;\  &
            $a/M=0.9$, $l/M^{2}=-5$;\  &
            $a/M=0.9$, $l/M^{2}=-1$;\  &
            $a/M=0.9$, $l/M^{2}=-0.5$ \\
            \includegraphics[width=0.25\textwidth]{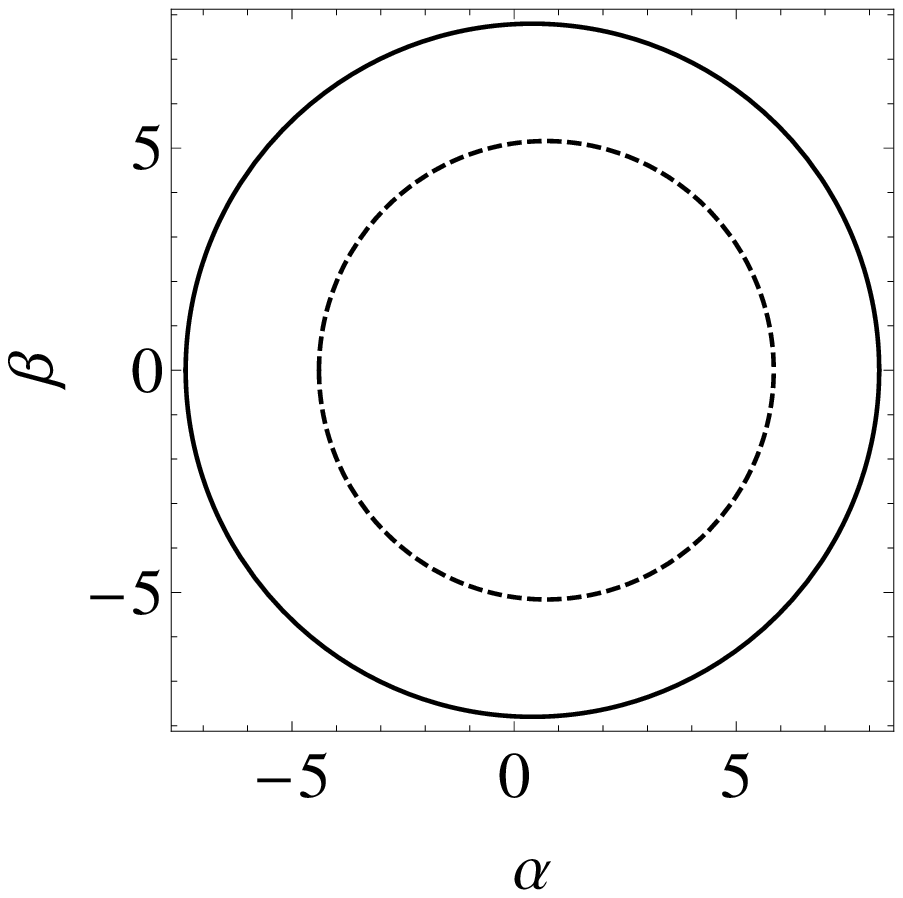} &
            \includegraphics[width=0.25\textwidth]{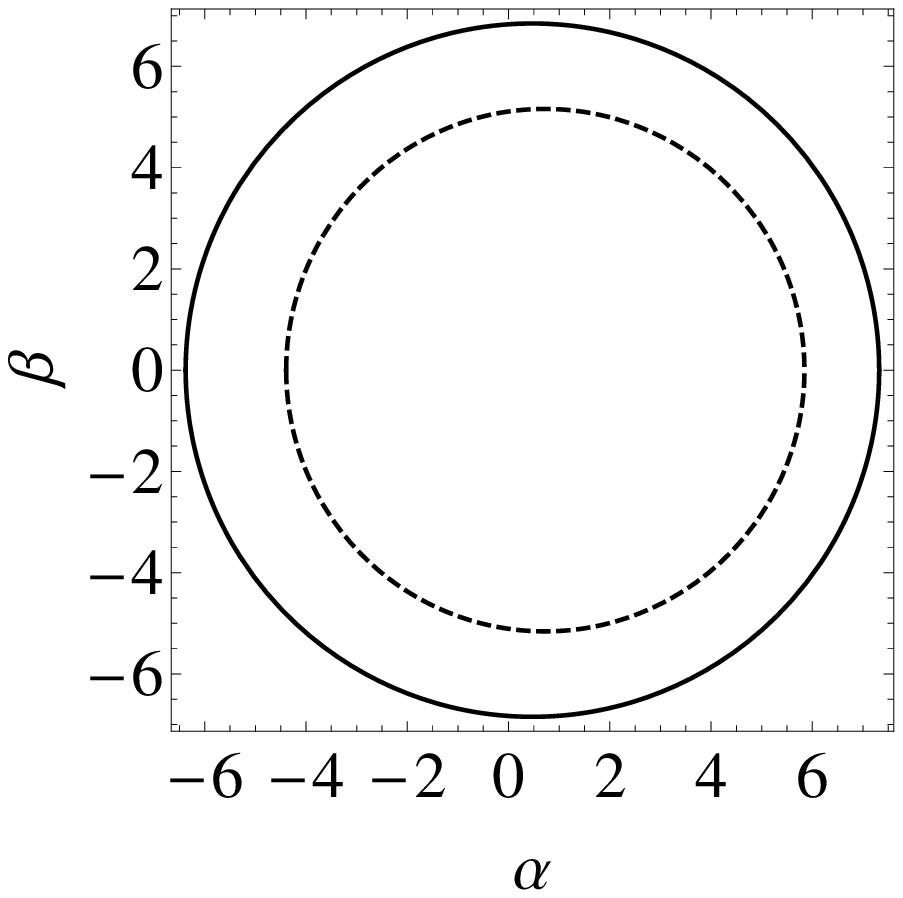} &
            \includegraphics[width=0.25\textwidth]{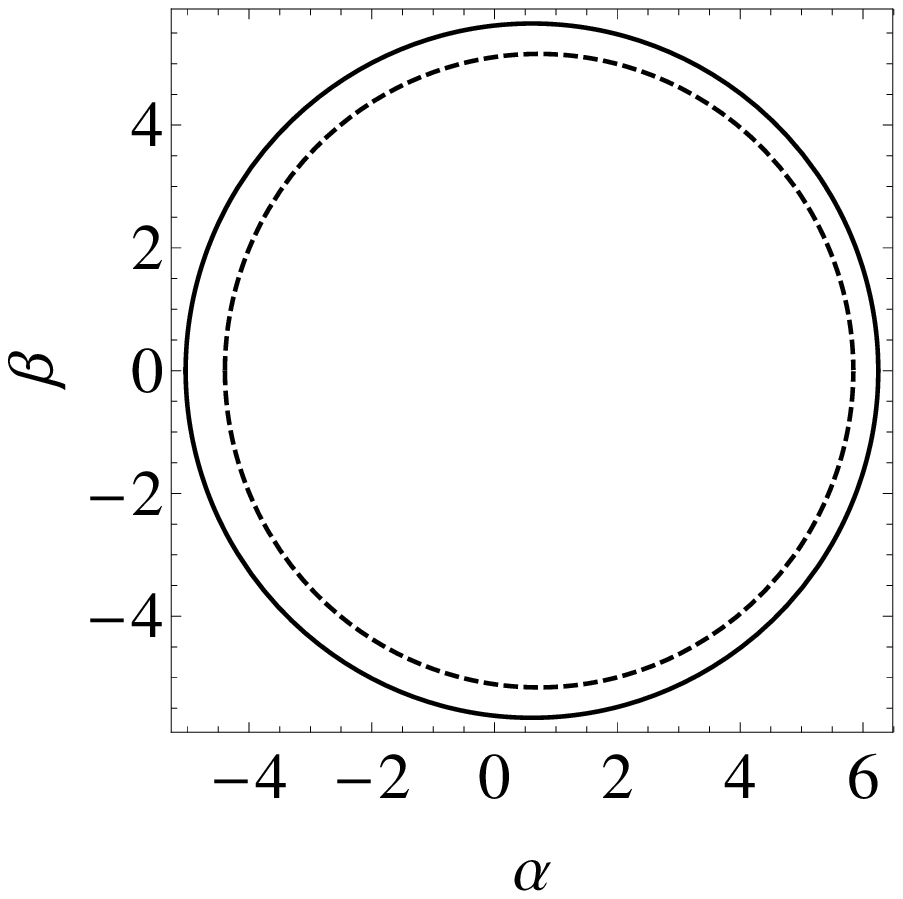} &
            \includegraphics[width=0.25\textwidth]{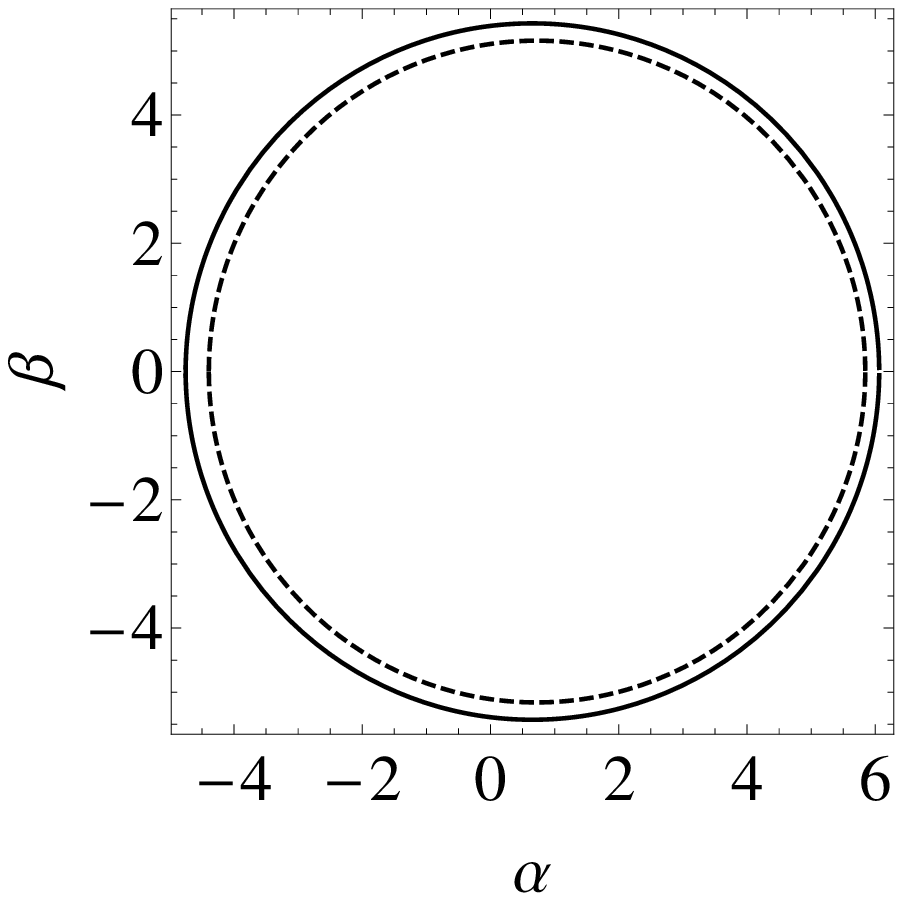} \\
            $a/M=0.5$, $l/M^{2}=-10$;\  &
            $a/M=0.5$, $l/M^{2}=-5$;\  &
            $a/M=0.5$, $l/M^{2}=-1$;\  &
            $a/M=0.5$, $l/M^{2}=-0.5$ \\
            \includegraphics[width=0.25\textwidth]{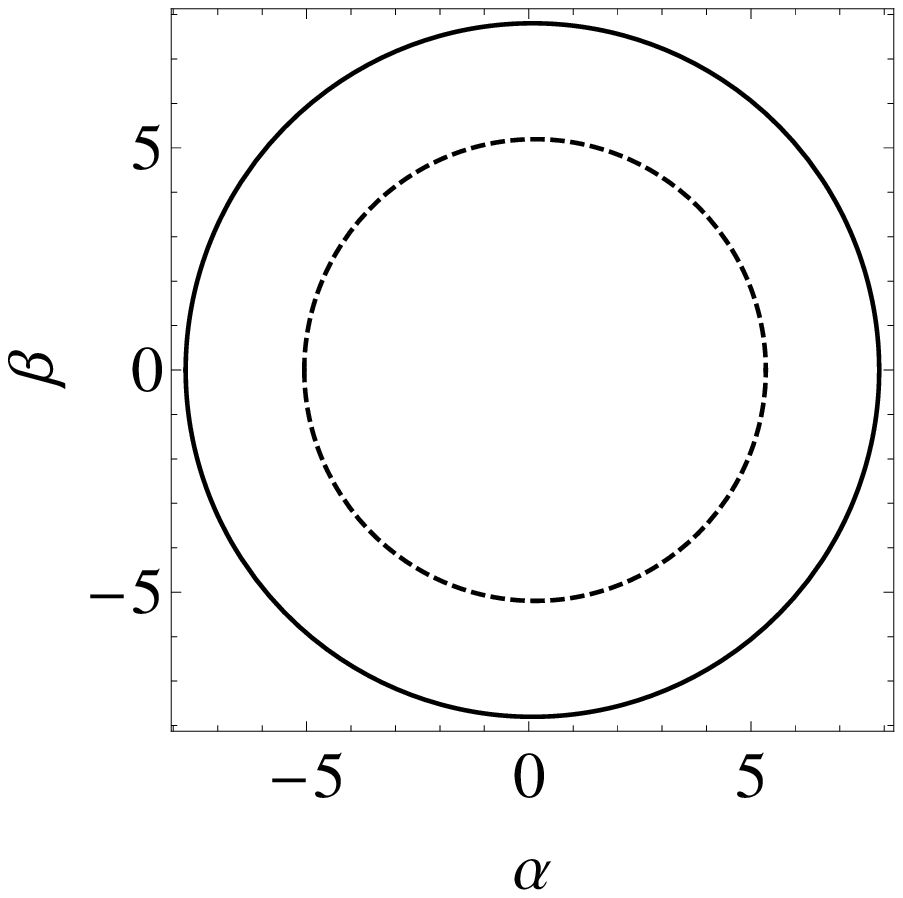} &
            \includegraphics[width=0.25\textwidth]{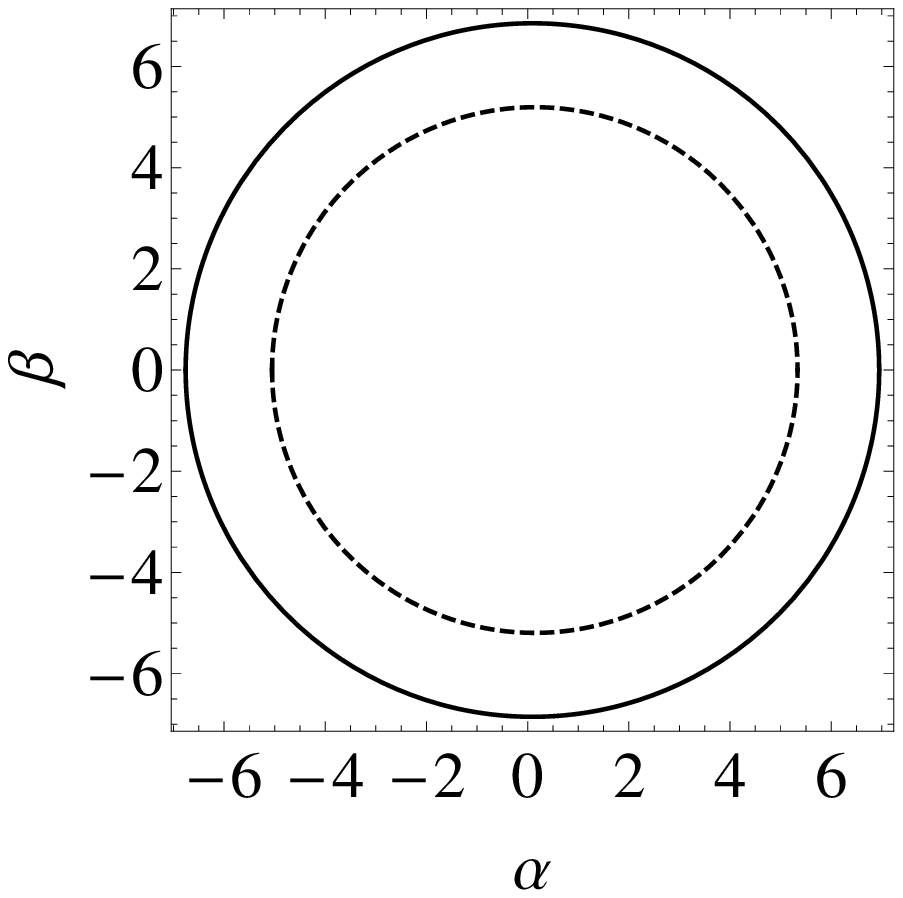} &
            \includegraphics[width=0.25\textwidth]{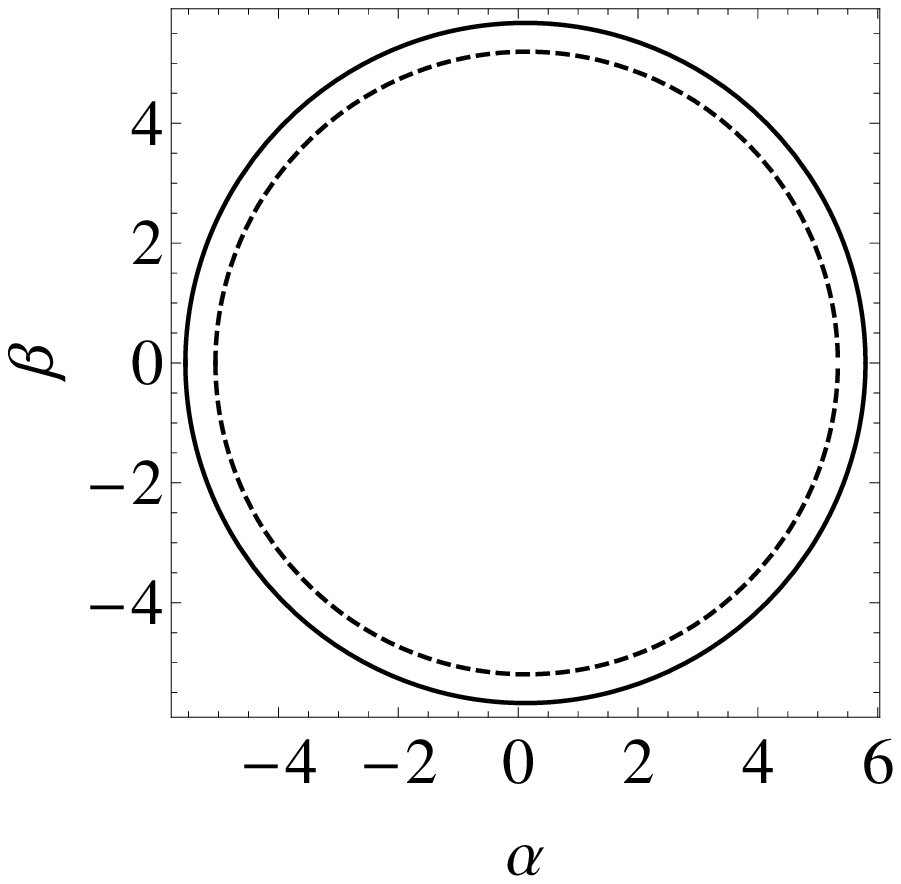} &
            \includegraphics[width=0.25\textwidth]{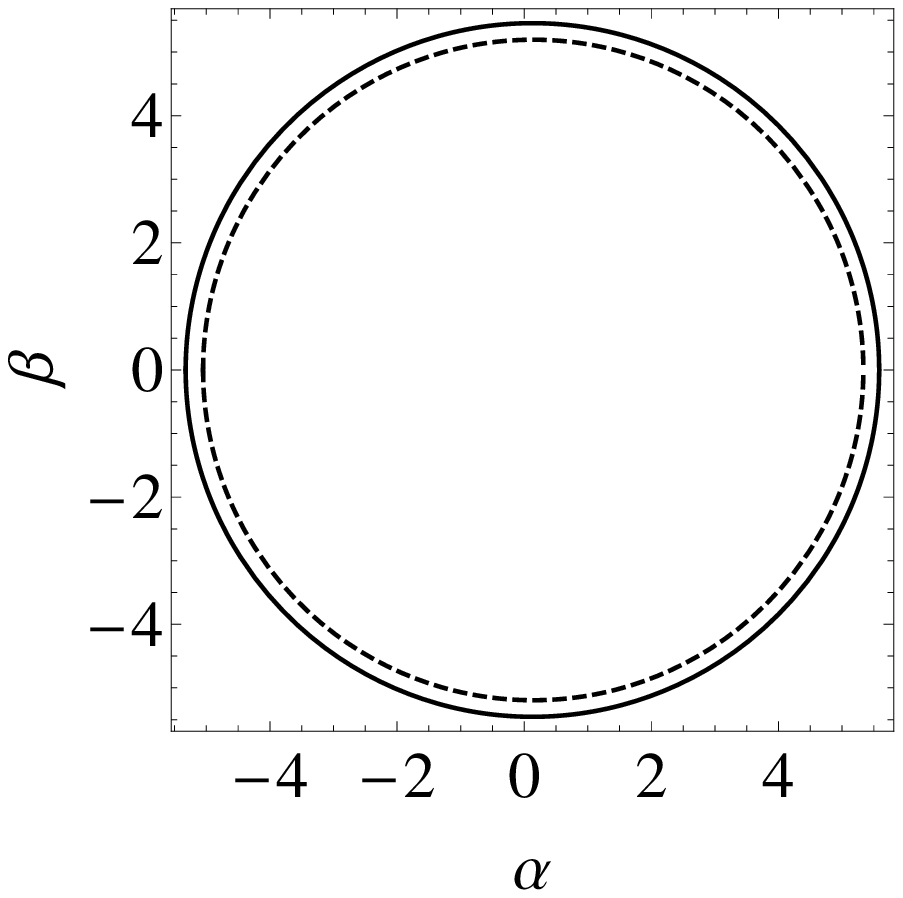} \\
            $a/M=0.1$, $l/M^{2}=-10$;\  &
            $a/M=0.1$, $l/M^{2}=-5$;\  &
            $a/M=0.1$, $l/M^{2}=-1$;\  &
            $a/M=0.1$, $l/M^{2}=-0.5$ \\
        \end{tabular}}
\caption{\footnotesize{The shadow of the generalized Kerr black hole in the model with function $\mu(r)=M e^{-l/r^{2}}$ (solid line) and the Kerr black hole (dashed line)
with inclination angle $\theta_{0}=\pi/4\ rad$ for different
values of parameters $a$ and $l$. The parameter $M$ of both solutions
is set equal to 1. The celestial coordinates $(\alpha,\beta)$ are
measured in the units of parameter $M$. } }
        \label{WS_a4minus}
\end{figure}

In Figs. \ref{WS_a3}-\ref{WS_a4minus} are presented the shadows of the Generalized Kerr black hole and its corresponding Kerr solution in case when $\mu(r)=Mexp(-l/r^2)$ for two inclination angles ($\pi/2$ and $\pi/4$ radians), and several values of the spin and the $l$ parameters.

Figs. \ref{WS_a5minus} and \ref{WS_a6minus} present the shadow of the Generalized Kerr black hole with $a>M$ as well as its corresponding Kerr naked singularity.

When the spin and the absolute value of the $l$-parameter are both small, the shadow of the Generalized Kerr black hole is very close to that of the Kerr black hole. Furthermore, when the parameter $l$ is positive, the shadow of the Generalized Kerr black hole is inside of its corresponding Kerr black hole shadow. This is not true in the opposite case.

In conclusion, we have shown that the presence of exotic matter can influence  the  geometry of black holes
and can leave observable imprints in the shadows of the black holes.

\end{document}